\documentclass[astrosymb, tighten, twocolumn]{aastex631}
\usepackage[utf8]{inputenc}
\usepackage{graphicx}
\usepackage{wrapfig}
\usepackage{amsmath}
\graphicspath{{./}{figures/}}

\usepackage{mathptmx,txfonts,tikz,bm}

\newcommand{\feh}{\ensuremath [\mathrm{Fe/H}]}
\newcommand{\amlt}{{\ensuremath \alpha_{\text{MLT}}}}

\renewcommand{\edit}[2]{{
\ifcase#1\or
#2
\or
\textbf{#2}
\else
#2
\fi%
}}

\begin{document}

\correspondingauthor{Christopher Lindsay}
\email{christopher.lindsay@yale.edu}

\title{Mixed Mode Asteroseismology of Red Giant Stars Through the Luminosity Bump}

\author[0000-0001-8722-1436]{Christopher J. Lindsay}
\author[0000-0001-7664-648X]{J. M. Joel Ong}
\author[0000-0002-6163-3472]{Sarbani Basu}
\affiliation{Department of Astronomy, Yale University, PO Box 208101, New Haven, CT 06520-8101, USA}

\shortauthors{Lindsay, Ong \& Basu}
\shorttitle{Asteroseismology of the Red Giant Branch Bump}

\begin{abstract}
Most current models of low mass red giant stars do not reproduce the observed position of the red giant branch luminosity bump, a diagnostic of the maximum extent of the convective envelope during the first dredge up. Global asteroseismic parameters, the large frequency separation and frequency of maximum oscillation power, measured for large samples of red giants, show that modeling convective overshoot below the convective envelope helps match the modeled luminosity bump positions to observations. However, these global parameters cannot be used to probe envelope overshoot in a star-by-star manner. Red giant mixed modes, which behave like acoustic modes at the surface and like gravity modes in the core, contain important information about the interior structure of the star, especially near the convective boundary. Therefore, these modes may be used to probe interior processes, such as overshoot. Using a grid of red giant models with varying mass, metallicity, surface gravity, overshoot treatment, and amount of envelope overshoot, we find that changing the overshoot amplitude (and prescription) of overshoot below the convection zone in red giant stellar models results in significant differences in the evolution of the models' dipole mixed-mode oscillation frequencies, the average mixed mode period spacing, $\langle \Delta P \rangle$, and gravity mode phase offset term, $\epsilon_g$. 
\end{abstract}

\keywords{asteroseismology - stars: solar-type - stars: oscillations - stars: interiors}

\section{Introduction} 
\label{sec:intro}

Improving the treatment of interior mixing beyond convective boundaries in stellar models is vital to many different areas of astrophysics. In particular, improvements to our understanding of mixing processes below the boundaries of convective envelopes in low mass red giant stars has the potential to improve models of red giants, which can, in turn, improve parameter estimations of mass, radius, and age for these stars. Due to their high luminosities, red giants can be observed with high signal-to-noise even from large distances and hence, red giants play an important role in the study of the Milky Way, since their ages are vital in constraining the chemo-dynamical evolution of the Galaxy.

The extent to which fusion products in red giants are brought to the surface depends on the properties of interior mixing below convective envelopes. Therefore, modelling inferences based on surface elemental abundances depend on good descriptions of interior mixing. For example, calibrations of stellar mass and age for low mass (between 1 and 2 M$_{\odot}$) red giant branch stars between the first dredge up and red giant branch luminosity bump are based on the observed carbon to nitrogen ratio, which in turn depends on the details of interior mixing of red giants \citep{Salaris2015}. Mixing at convection zone boundaries also affects the differential rotation profile \citep{Pinsonneault1997} and rate of lithium depletion \citep{Baraffe2017}. 

Many methods can be used to investigate the extent of mixing past convective boundaries. Investigations using 2D and 3D hydrodynamic simulations of regions of convective overshoot can inform how overshooting should be treated in 1 dimensional stellar models \citep[e.g.][]{Pratt2017, Korre2021, Anders2021}. Statistical studies comparing 1D stellar models with varying amounts of overshoot to ensemble observations of red giants can also be used to constrain interior mixing. The extent of mixing beyond the convective boundary has been linked to a change in position of the red giant branch luminosity bump or RGBb \citep{Alongi1991}, a diagnostic of the maximum extent of the convective envelope during the first dredge up stage of stellar evolution. \citet{Khan2018} compared stellar models of red giant stars with differing amounts of mixing past the convective envelope boundary to ensemble observations of red giants from \citet{APOKASC1} and were able to place boundaries on the amount of overmixing based on the observed and modeled locations of the RGBb. Since the oscillations of these stars encode information about the stellar interiors, another method of constraining the amount of mixing below convective envelope boundaries is to look at the influence of overshoot on red giant oscillations. 

Space based photometry missions, with their long temporal baselines and high frequency resolution, now provide the means to perform measurements of individual oscillation modes of red giant stars. To date, almost all asteroseismic studies of red giants have relied on the average seismic parameters, $\Delta \nu$, the large frequency separation, and $\nu_{\text{max}}$, the frequency of maximum oscillation power. These global asteroseismic variables are used in conjunction with metallicity and effective temperature data to determine masses and ages for large samples of stars \citep[e.g.][]{APOKASC1,APOKASC2}. These have been very useful both directly, in the study of the Milky Way and Galactic archaeology \citep[][etc.]{Nidever2014,Valentini2019,Lian2020,Spitoni2020,Hon2021,Zinn2021}, as well as indirectly, as training sets for machine-learning based methods to derive more accurate red giant stellar properties from spectra \citep[e.g.][]{Ness2015,Martig2016,Ting2021}. Average spacings in the periods of gravity waves propagating in red giants stars have also proven useful for distinguishing between inert helium core red giant branch stars and core helium burning red clump stars \citep[][]{Bedding2011}. However, these studies of global parameters cannot directly constrain the internal physical processes taking place in red giants on a star by star level. 

The Sun, being our closest star, provides a useful laboratory for studying how overshoot below convective envelopes influences the oscillation modes of stars. Substantial overshoot below the convection zone, modeled as a deepening of the adiabatic temperature gradient of the convection zone (penetrative overshooting, see \autoref{sec:modelling}), will cause a sound speed discontinuity at the solar envelope position. This acoustic glitch is detectable using helioseismic observations of p-mode oscillatory signatures \citep{Gough1993}. Using p-mode frequencies of the Sun, the amount of overshoot below the solar convection zone base was found to be less than $0.03$ pressure scale heights ($H_p$) \citep{Basu1994, Basu1997}, though changing the prescription of overshoot can raise the limit to $0.07$ $H_p$ \citep{Monteiro1994, Dalsgaard1995}. Today, there is a wealth of seismic data for many red giants available from missions such as \textit{Kepler} \citep{Kepler_inst}, and \textit{TESS} \citep{TESS_inst}, and this seismic data may be able to constrain overshoot in these stars.

For some high mass stars on the main sequence, individual mode frequencies have been used to look at the effects of overshooting above convective cores. These analyses have shown that core overshoot in intermediate and high mass stars is needed to match observed frequencies from \textit{Kepler} \citep[e.g.][etc.]{Moravveji2016, Pedersen2018, Michielsen2019}. Core overshoot is important for determining the ages of evolved stars as well, since the amount of core overshoot will dictate when a star evolves off the main sequence and therefore can determine a star's main sequence lifetime. On the other hand, envelope overshoot below a star's convective envelope will change the course of a star's evolution after it leaves the main sequence, and that is the focus of this paper. As we show in \autoref{sec:asteroseismic}, implementing envelope overshoot in low mass red giant stellar models alters their interior structure and evolution up the red giant branch. 

Our goal is to study how different prescriptions and amplitudes of overshoot below convective envelopes affect the individual oscillation mode frequencies in low mass red giant stars. Another important question is to determine whether real asteroseismic data for specific stars can be used to determine their internal overshooting properties. To do this, we use a grid of stellar models with varying mass, metallicity, and convective overshoot prescription (detailed in \autoref{sec:modelling}) calculated using the 1-D stellar evolution code MESA \citep{Paxton2011, Paxton2013, Paxton2015, Paxton2018, Paxton2019}. The different initial conditions of these stellar models will result in varying asteroseismic properties between them.

\begin{figure*}[ht!]
    \centering
    \includegraphics[scale=0.38]{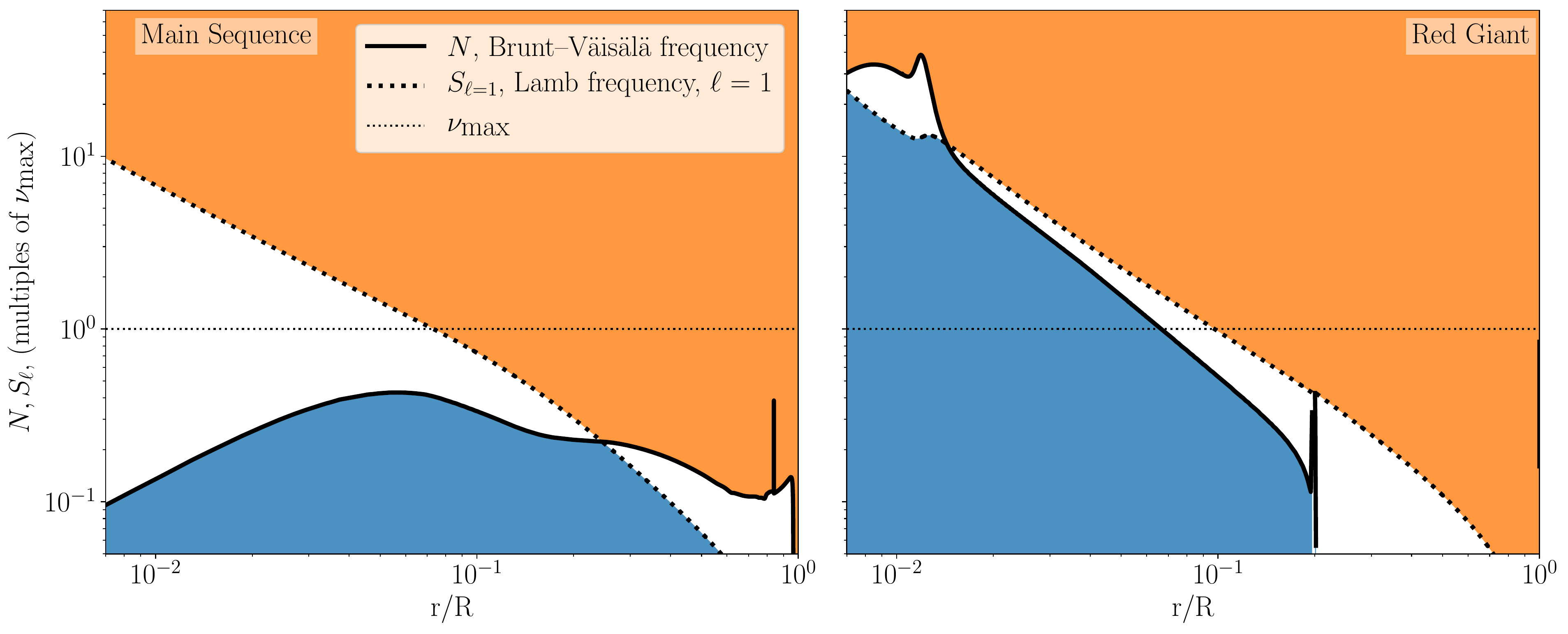}
    \caption{Propagation diagrams of a main sequence star (left) and a red giant branch star (right) with the Brunt–Väisälä frequency profile indicated with the solid lines and the $\ell=1$ Lamb frequency profiles indicated with the dotted lines. The horizontal line shows the frequency of $\nu_{\text{max}}$. Mixed modes with frequency near $\nu_{\text{max}}$ are excited in red giant branch stars, since the g-mode propagation cavity (blue) is not well separated from the p-mode propagation cavity (orange). In regions of a star where the Brunt–Väisälä frequency is imaginary, the stellar fluid is unstable to convection and there is a convection zone. }
    
    \label{fig:mixed_mode_explaination}
\end{figure*}

Stars with convective envelopes, including red giants, oscillate with multiple modes excited by convection \citep{Goldreich1977a,Goldreich1977b}. Such oscillations are usually referred to as solar-like oscillations, since the Sun exhibits them as well. The oscillation power spectrum of a solar-like oscillator has a bell-shaped envelope whose maximum frequency position is at $\nu_{\text{max}}$, the frequency of maximum oscillation. Empirically, $\nu_{\text{max}}$ scales as approximately $g/\sqrt{T_{\text{eff}}} \text{ or } M/(R^2\sqrt{T_{\text{eff}}})$ where $g$ is the surface gravity, $M$ is the mass, $R$ is the radius, and $T_{\text{eff}}$ is the effective temperature \citep{Kjeldsen1995}.

The long temporal baselines from \textit{Kepler} \citep{Kepler_inst} and from the TESS \citep{TESS_inst} continuous viewing zones (CVZs) allow for the resolution of individual mode frequencies in the power spectra of thousands of red giant stars; these individual modes allow us to probe the internal structure of these stars. Evolved stars in particular support the propagation of two different kinds of waves at frequencies close to $\nu_\text{max}$, governed by two characteristic frequencies: the Lamb frequency (dotted curves in \autoref{fig:mixed_mode_explaination}), and the Brunt–Väisälä (or buoyancy) frequency (solid curves). Waves where the restoring force is pressure can only propagate at frequencies above both of these quantities (within the orange regions in \autoref{fig:mixed_mode_explaination}); normal modes in this regime are called p modes. For any given degree $\ell$, pressure modes of consecutive radial orders $n$ are approximately equidistant in frequency, and this separation is called the ``large frequency separation'', $\Delta \nu$. The average large frequency separation scales approximately with the square root of the mean density of the star ($\Delta \nu \propto \sqrt{M/R^3}$). Conversely, waves where the restoring force is buoyancy are called gravity modes, or g-modes, and have frequencies below both the Lamb and Brunt–Väisälä frequencies (blue regions in \autoref{fig:mixed_mode_explaination}).

In main sequence stars, the p-mode and g-mode cavities are well separated both spatially and in frequency, so the normal modes supported by their structures with frequencies around $\nu_{\text{max}}$ are purely acoustic. As these stars evolve up the red giant branch, due to the large difference in density between the red giant core and envelope, oscillation modes of red giant stars will exist in two different regions: the core, which supports g-modes, and the convective envelope, which supports p-modes (see \autoref{fig:mixed_mode_explaination}). Unlike in main sequence or sub-giants, the resonant cavities in red giants are coupled by an intermediate region. So-called mixed modes with frequencies near $\nu_{\text{max}}$ arise due to mode coupling across this evanescent region, as described in \citep{Osaki1975, Aizenman1977, Shibahashi1979, Takata_2016a}. Red giant mixed modes are p-like (i.e., acoustic modes) at the surface and g-like (i.e., gravity modes) in the core. The boundary between the p-mode and g-mode cavities across which mode coupling occurs lies near the base of the red giant convection zone, so mixed modes sample the interior structure of these star near the convective boundary. Thus, the mixed mode frequencies can be used to study overshooting processes. See \citet{2017A&ARv..25....1H}, and references therein, for an extensive review of giant star seismology. 

The frequency separations between adjacent p-modes deviate from the average value of $\Delta \nu$, though these deviations are small. The deviations are accounted for by expressing p-mode frequencies as $\nu_{\text{n}, \ell} = \Delta \nu(n + \ell/2 + \epsilon_p)$, where $\epsilon_p$ is a phase factor and can be regarded as the offset of the radial modes from 0 frequency. Analogous to how p-modes are approximately equidistant in frequency, g-modes are approximately equally spaced in period with the period spacing between adjacent g-modes approximately given by, 
\begin{equation}
    \Delta \Pi_0 = 2 \pi^2 \bigg(\int_{r_1}^{r_2} N(r) \text{d}r/r\bigg)^{-1}\label{eq:deltapi}
\end{equation}
where $r_1$ and $r_2$ are the boundaries of the radiative zone where g-modes can propagate. For modes of a given $\ell$, the approximate period spacing is given by $\Delta \Pi_{\ell} = \Delta \Pi_0 /\sqrt{\ell(\ell+1)}$ \citep{Tassoul1980}. g-mode periods can then be expressed as $\Pi(n_g, \ell) = \Delta \Pi_{\ell}(n_g  + \epsilon_g)$ \citep{Aizenman1977,Tassoul1980,Aerts-JCD-Kurtz2010} where again $\epsilon_g$ is a phase or offset term. As we will show, this gravity-mode phase offset has diagnostic potential when it comes to studying envelope overshoot.

First, \autoref{sec:modelling} goes over our different modelling prescriptions for convective overshoot from the envelope and the construction of our grid of models. In \autoref{sec:RGBb} we discuss how different overshooting prescriptions implemented in red giant stellar models change their evolution up the red giant branch and we show in \autoref{sec:asteroseismic} how different overshooting prescriptions change the interior structure, individual mixed mode oscillations, and asteroseismic properties of our grid of red giant models. Prospects for studying the asteroseismic differences between different overshoot prescriptions in real seismic data is discussed in \autoref{sec:dis} and we conclude in \autoref{sec:conclusion}.

\section{Modelling Procedure}
\label{sec:modelling}
Different treatments of overshoot change the interior structure and oscillation modes of red giant stellar models. Convective overshoot occurs when parcels of fluid move past convection zone boundaries into radiative zones, and is currently modelled in a number of different ways in 1D stellar evolution codes. In our modeling, we consider two distinct ways of treating the temperature gradient at the mixing boundary. One, where the overshooting region has an adiabatic temperature gradient and another, which assumes that the overshooting region has a radiative temperature gradient. The prescription that assumes an adiabatic temperature gradient is historically called ``penetrative convection'' or ``penetrative overshoot'' while the prescription that assumes a radiative temperature gradient is generally called ``overshooting'' or ``overmixing'' \citep{Zahn1991}. 

In addition, we consider two different profiles for the mixing coefficient within the overshooting region below the convective envelope. The first profile we use is the  so-called step overshoot, where the high mixing coefficient of the convection zone is extended into the radiative zone by the overshooting region depth, defined to be $f_{\text{ov, step}}H_p$ or $f_{\text{ov, step}}r_\text{cz}$, whichever is smaller. Here $f_{\text{ov, step}}$, is a free parameter, $H_p$ is the pressure scale height at the convective boundary, and $r_{cz}$ is the radius of the convection zone. $f_{\text{ov, step}}$ is often called $\alpha_{\text{ov}}$ in the literature, but in this work we will use $f_{\text{ov, step}}$ as the step overshooting parameter. Step profile overshoot has traditionally been used in stellar modelling and recent work suggests that step overshoot best fits the results of convection simulations \citep[][]{Anders2021}. Regardless of whether step penetrative overshoot or step overmixing is implemented, the composition is uniform and the same as in the convection zone. 

The second profile we used when modelling overshoot, exponential overshoot, was first formulated by \citet{SchlattlandWeiss1999} and is based on the convection simulations of \citet{Blocker1998} and \citet{Freytag1996}. Exponential overshoot was originally composed to describe overshooting from convective cores, but we implement the same profile shape below red giant convective envelopes. The exponential profile prescription assumes that overshoot is a diffusive process with diffusion coefficient:
\begin{equation}
    D_{\text{ov}} = D_{\text{conv}} \exp \bigg( -\frac{2 z}{f_{\text{ov, exp}} H_p} \bigg).
    \label{eq:exponential_diffusion}
\end{equation}
Here, $f_{\text{ov, exp}}$ is an analogue to $\alpha_{\text{ov}} \text{ or } f_{\text{ov, step}}$ in the step overshoot prescription and is a free parameter. The diffusion coefficient under the convective boundary in the radiative zone, $D_{\text{ov}}$, is calculated with equation 2 using the overshoot parameter, $f_{\text{ov, exp}}$, the pressure scale height, $H_p$, the geometric distance from the edge of the convection zone into the radiative zone, $z$, and the diffusion coefficient calculate by MESA just inside the convection zone from the boundary, $D_{\text{conv}}$. Again, there are two different cases within exponential overshoot, one where the temperature gradient beyond the convection zone is assumed to be radiative and another where the temperature gradient changes from adiabatic to radiative with the same exponential profile drop off as the diffusion coefficient.

Given that the nature of overshoot from convection is not known, its treatment in stellar models is often done under several assumptions. How exactly 1D stellar modelling codes should treat overshoot in different stars is unknown. In order to investigate how the aforementioned different overshoot prescriptions detailed change the asteroseismic properties of red giant stars, we calculate 4 sets of red giant models using MESA version r12778 \citep{Paxton2011, Paxton2013, Paxton2015, Paxton2018, Paxton2019}, using an Eddington gray atmospheric boundary condition, the mixing-length prescription of \citet{CoxGiuli1968}, and the diffusion of heavy elements as in \citet{Burgers1969}. We evolve models with the aforementioned physics, incorporating prescriptions for convective envelope overshooting, as follows: 

\begin{enumerate}
    \item Step Overmixing: we calculate step overshoot mixing from the base of the red giant's convective envelope into the radiative core. The high mixing coefficient of the convection zone is extended into the radiative zone by the overshooting region depth, defined to be $f_{\text{ov, step}}H_p$.
    \item Exponential Overmixing: we calculated exponential overshoot mixing from the base of the convective envelope into the radiative core. The diffusion coefficient below the convection zone is defined by \autoref{eq:exponential_diffusion}. 
    \item Step Penetrative Overshoot: we calculate step overmixing from the convective envelope into the core, with an adiabatic temperature gradient in the overshoot region (reaching a distance of $f_{\text{ov, step}}H_p$ into the radiative core from the convective boundary). 
    \item Exponential Penetrative Overshoot: we calculate exponential overmixing from the convective envelope into the core plus an adjustment to the temperature gradient ($\nabla$) profile such that $\nabla = f \nabla_{ad}$ below the convective envelope with the adiabatic temperature gradient $\nabla_{ad}$ and factor $f = \exp\big( \frac{-2 z}{f_{ov, exp} H_p}\big)$
\end{enumerate}

For each overshoot prescription, we evolve stellar models with masses $M/M_{\odot} \in [1.0, 2.0]$ in steps of $0.2 M_{\odot}$ and with solar calibrated values of initial helium abundance, $Y_0 = 0.272$, initial metallicity, $Z_0 = 0.0186$, and mixing length parameter, $\alpha_{\text{MLT}} = 2.08$. For the $1.4 M_{\odot}$ mass, we also evolve stellar models with $\feh = -0.2$ and 0.2 to investigate metallicity dependence. In order to densely sample the region of red giant branch evolution around the RGBb, we output stellar models in log($g$) steps of 0.025 from log($g$) = 3.5 to log($g$) = 2.25, a total of 51 outputted models per stellar evolutionary track. 

For both step overmixing and step penetrative overshoot, we use varying values of $f_{\text{ov, step}}$ ranging from 0.0 (no overshoot) to 0.6 in steps of 0.1. For the exponential overmixing and exponential penetrative overshoot cases, we evolve red giant models with values of $f_\text{{ov, exp}}$ ranging from 0.0 to 0.06 in steps of 0.01. We choose these ranges based on matching the modelled location of the bump to the over density of stars in the APOKASC \citep{APOKASC1} catalog (see \autoref{fig:RGBb_overshoots} based on \citet[][]{Khan2018}). The parameters of our outputted red giant models are summarized in \autoref{table:params}.

\begin{table}[ht!]
\centering
\begin{tabular}{ccc}
\hline
\multicolumn{1}{|c|}{Quantity} & \multicolumn{1}{c|}{Range} & \multicolumn{1}{c|}{Step Size} \\ \hline
Mass [$M_{\odot}$]                          & \{1.0, 2.0\}               & 0.20                            \\
Metallicity [Fe/H]                   & \{-0.2, +0.2\}             & 0.20                            \\
$f_{\text{ov, exp}}$           & \{0.0, 0.04\}              & 0.01                           \\
$f_{\text{ov, step}}$          & \{0.0, 0.4\}               & 0.10                            \\
log($g$)                       & \{3.5, 2.25\}              & 0.025                          \\    
$\alpha_{\text{MLT}}$          & 2.08                       &                                \\    
\end{tabular}
\caption{Summary of our red giant model grid parameters. The higher and lower metallicity red giant model tracks were only calculated for the 1.4 $M_{\odot}$ mass.}
\label{table:params}
\end{table}

Since element diffusion is known to produce unphysical results for high-mass stars with very thin or non-existent, convective envelopes, we implement the soft diffusion cut off proposed in \citet{Viani_2018} by multiplying the diffusion coefficient by a mass dependent factor $F$ given by,

\begin{equation}
F = \left\{
        \begin{array}{ll}
            \exp\bigg[ - \frac{(M/M_{\odot} -1.25)^2}{2(0.085)^2}\bigg] & \quad M/M_{\odot}>1.25 \\
            1 & \quad M/M_{\odot}\leq 1.25
        \end{array}
    \right.
\end{equation}

For each model in our grid of stellar models, we use the stellar oscillation code GYRE (version 6.0 \citep{Townsend2013}) to calculate the radial ($\ell = 0$) mode frequencies as well as the dipole ($\ell = 1$) mixed mode oscillation frequencies.

\section{The Effects of Overshoot on Red Giant Tracks}
\label{sec:RGBb}
\subsection{The Red Giant Branch Luminosity Bump}

The red giant branch luminosity bump or RGBb is a part of the ascending red giant branch, characterized by a brief decrease in luminosity associated with the star's hydrogen-burning shell advancing to and passing the discontinuity in hydrogen abundance left behind by the first dredge up \citep[see][and references therein]{2015MNRAS.453..666C}. The decrease and subsequent increase in luminosity means that stars will spend a little longer in that section of the Hertzsprung-Russell diagram, leading to an accumulation of stars in that region when looking at ensemble studies of stellar clusters. Most models of red giant stars without envelope overshoot do not accurately reproduce the observed location of the red giant branch luminosity \citep{Khan2018}.

\begin{figure*}[ht!]
    \centering
    \includegraphics[scale=0.45]{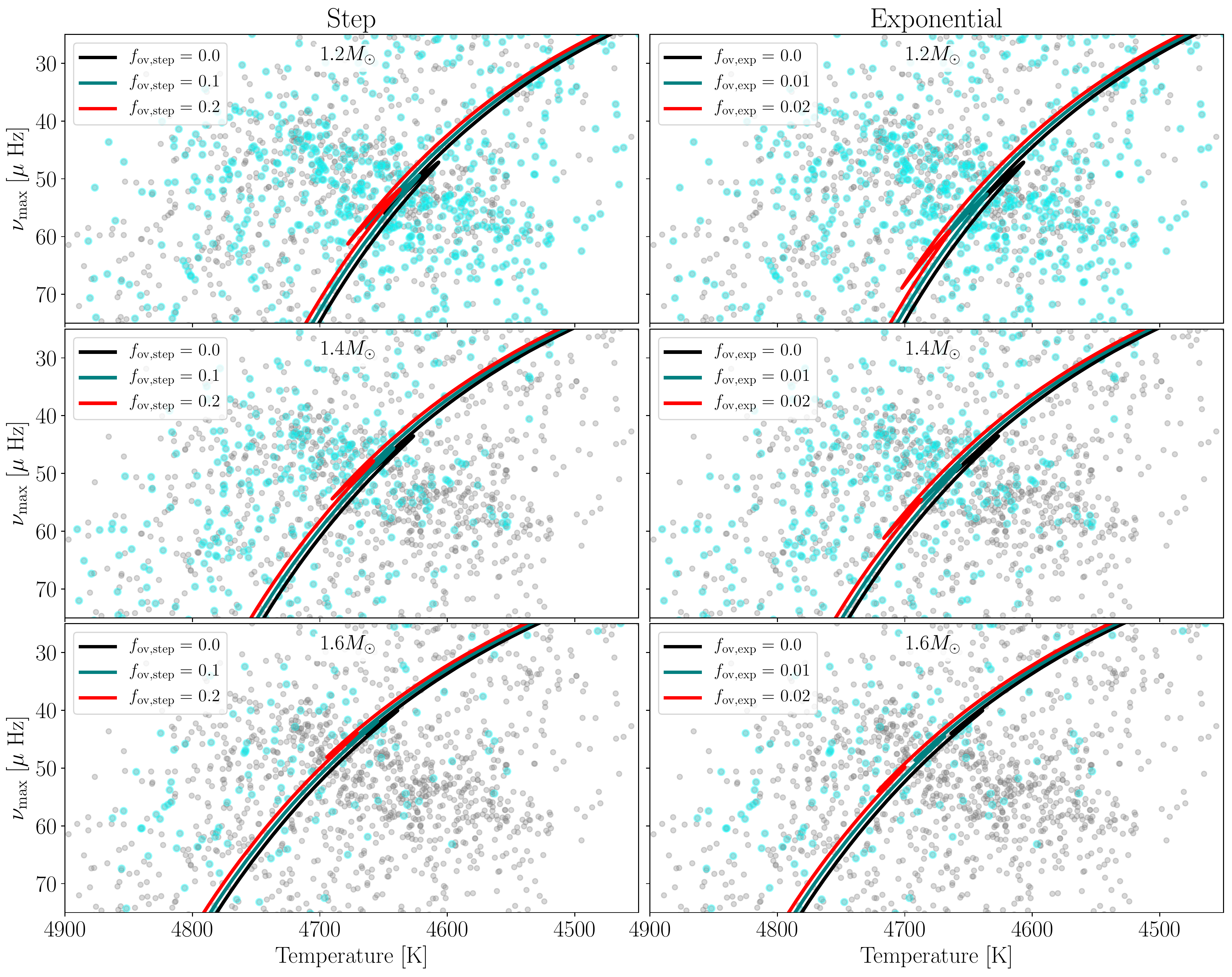}
    \caption{$\nu_{\text{max}}$ versus temperature evolutionary tracks for red giant models incorporating different amounts of step overmixing (left column) and exponential overmixing (right column). Evolutionary tracks with increasing overshoot amplitudes ($f_{\text{ov, step/exp}}$) are offset in temperature by 5 K from each other to make determining the location of the RGBb clearer. The gray points show the $\nu_{\text{max}}$ versus temperature positions of the stars in the red giant sample of the APOKASC catalog \citep{APOKASC1}. These points show the observed position of the RGBb and can be used to determine the amount of overshoot needed in models to match the data. Stars with seismic masses within 0.1 $M_{\odot}$ of the mass of the stellar model tracks are shown with cyan points.}
    \label{fig:RGBb_overshoots}
\end{figure*}

\citet{Khan2018} used the detection of the RGBb in stars observed by \textit{Kepler} that had temperature and metallicity values derived from APOGEE spectra \citep{2017ApJS..233...25A}, and noted that the RGBb location in the global asteroseismic variables, $\nu_{\text{max}}$ and $\Delta \nu$, show trends with mass and metallicity in line with expectations from models. However, models of varying mass and metallicity without overshooting from the convective envelope boundary do not agree with the RGBb position derived from observations; as we can verify \citep[see][Figure 1]{Khan2018}. Therefore, substantial overshooting from the base of the convective envelope, with a dependence on stellar mass and metallicity, must be considered for reproducing the location of the RGBb. We find that depending on stellar mass and metallicity, inserting step or exponential envelope overshoot of differing strengths into stellar models of red giants puts the modeled position of the RGBb bump more in line with the observed over density of red giant stars in the APOKASC catalog (see \autoref{fig:RGBb_overshoots}).

In \autoref{fig:RGBb_overshoots} we show that different values of $f_{\text{ov, step/exp}}$ changes the position of the RGBb. Matching the position of the modeled RGBb to the overdensity of red giant points in \autoref{fig:RGBb_overshoots} shows that for red giant models, different amplitudes of step or envelope overshoot can line up the modeled RGBb with observations, informing how much envelope overshoot should be used in stellar modelling \citep{Khan2018}. However, this analysis of populations of stars does not necessarily inform us of internal processes happening in individual red giants like convective overshoot since they focus on the global asteroseismic variables, $\nu_{\text{max}}$ and $\Delta \nu$. By employing the asteroseismic analysis techniques from \autoref{sec:intro} to our grid of red giant models in \autoref{sec:modelling}, we improve previous studies by investigating how different overshoot prescriptions change red giant interior structure and individual oscillation frequencies.

\subsection{Converting Between Step and Exponential Overshoot Parameters}
\begin{figure*}[ht!]
    \centering
    \includegraphics[scale=0.5]{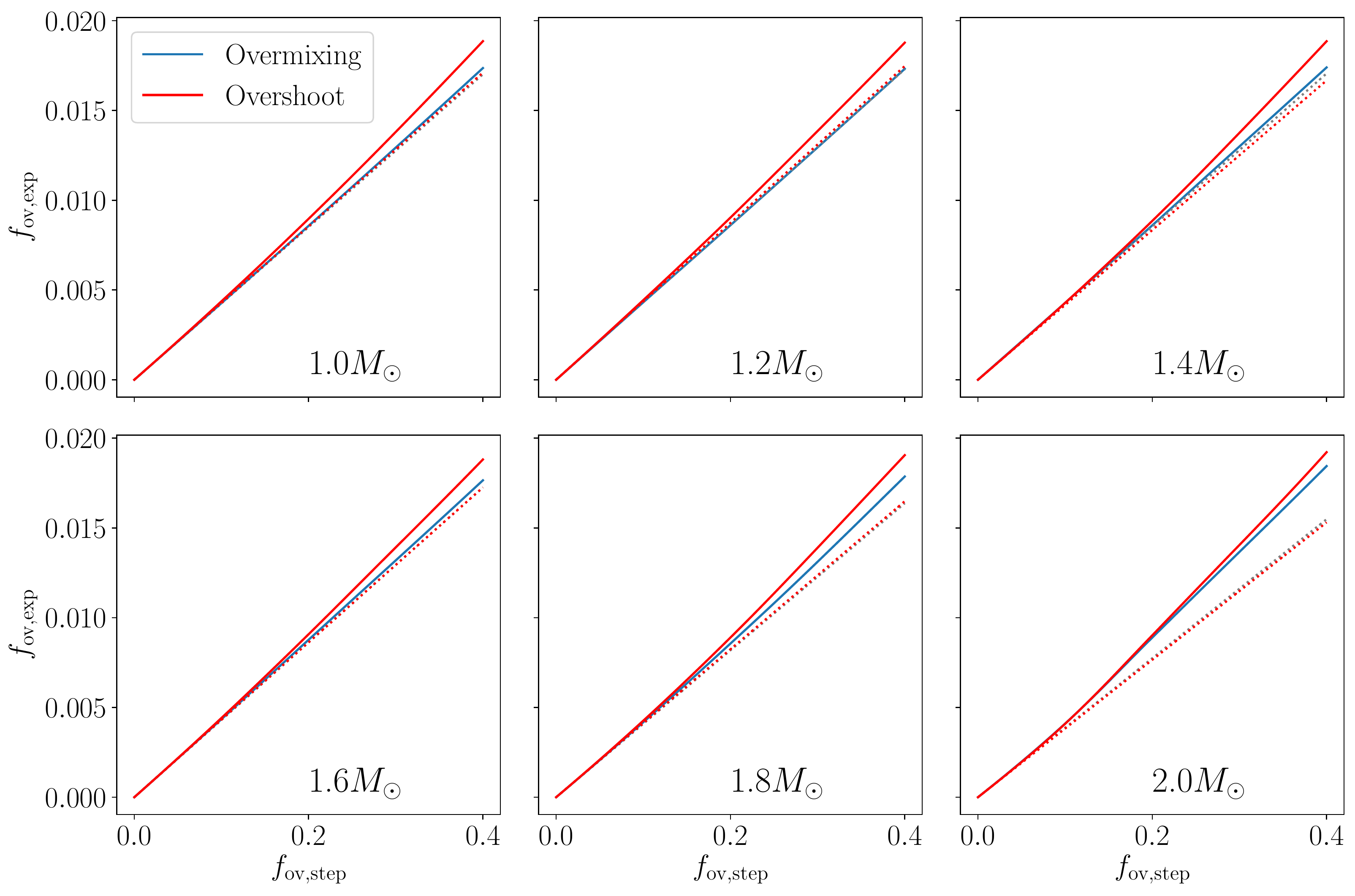}
    \caption{Calibration of mixing parameters $f_\text{ov}$ between different computational prescriptions for convective boundary mixing, using the RGBb as a spectroscopic anchor point. For each stellar mass, each solid curve shows the value of $f_\text{ov,exp}$ describing convective boundary mixing (CBM) with an exponential profile on the vertical axis, required to produce the same change in the position of the RGBb on the Kiel diagram as would be effected by the corresponding value of $f_\text{ov,step}$ describing CBM with a step-function profile on the horizontal axis. We anchor these spectroscopic values to the epoch where the luminosity of the stellar models along the red-giant evolutionary track first starts to decrease over time, during the first ascent up the RGB. The dotted lines in each panel are tangent lines to the curves of the same color at $f_{\text{ov, step/exp}} = 0$, where the two prescriptions must agree; we see that the calibrated relations exhibit increasing curvature (i.e. become increasingly nonlinear) with increasing stellar mass.}
    \label{fig:f_ov_calibration}
\end{figure*}

While we have seen that increasing $f_\text{ov, step/exp}$ shifts the RGBb position to higher values of $\nu_{\text{max}}$ and therefore higher values of $\log(g)$, step and exponential overshooting does so by different amounts. In order to properly compare the evolution of red giant models with different overshoot region shapes, we must first determine some way to convert between $f_{\text{ov, step}}$ and $f_{\text{ov, exp}}$. We do this by matching the position of the RGBb between evolutionary tracks in the Kiel diagram (i.e. $\log (g)$ vs. $T_{\text{eff}}$ space). We define the effective position of the RGBb for an evolutionary track to be the point on this diagram where the luminosities of the models first stop increasing over time during their initial ascent up the red giant branch. For each mixing prescription, this yields $\log (g)$ as a function of $f_\text{ov, step/exp}$; in practice, we find that this relation is bijective and equivalently specifies $f_\text{ov, step/exp}$ as a function of $\log (g)$ (at the loci of possible positions of the RGBb). For two different overshoot prescriptions, this yields pairs of $f_{\text{ov, step}}$ and $f_{\text{ov, exp}}$ for each prescription, each displacing the RGBb to the same value of $\log (g)$. This serves as the basis of our comparison between overshoot profile shapes. In \autoref{fig:f_ov_calibration} we plot $f_{\text{ov, exp}}$ against $f_{\text{ov, step}}$ for different masses, as calibrated by this procedure. We find that $f_{\text{ov, step}} \approx 20 f_{\text{ov, exp}}$ for our models of solar calibrated abundances within a mass range of $1.0 M_{\odot}$ to $2.0 M_{\odot}$ using both the overmixing and penetrative overshoot prescriptions. This relationship remains the same when using the middle or bottom position of the RGBb as a spectroscopic anchor point. We also note there is some curvature in the relationship between $f_{\text{ov, exp}}$ and $f_{\text{ov, step}}$ and that curvature increases slightly for higher mass models. However, such curvature is small within the range of values which we consider, and we neglect its effects when comparing $f_{\text{ov, exp}}$ and $f_{\text{ov, step}}$. We will use this conversion factor when making comparisons between step and exponential overshooting in our subsequent discussion.

\section{Results}
\label{sec:asteroseismic}

\subsection{Effects of Overshooting on Structure}

\begin{figure*}[ht!]
    \centering
    \includegraphics[scale=0.45]{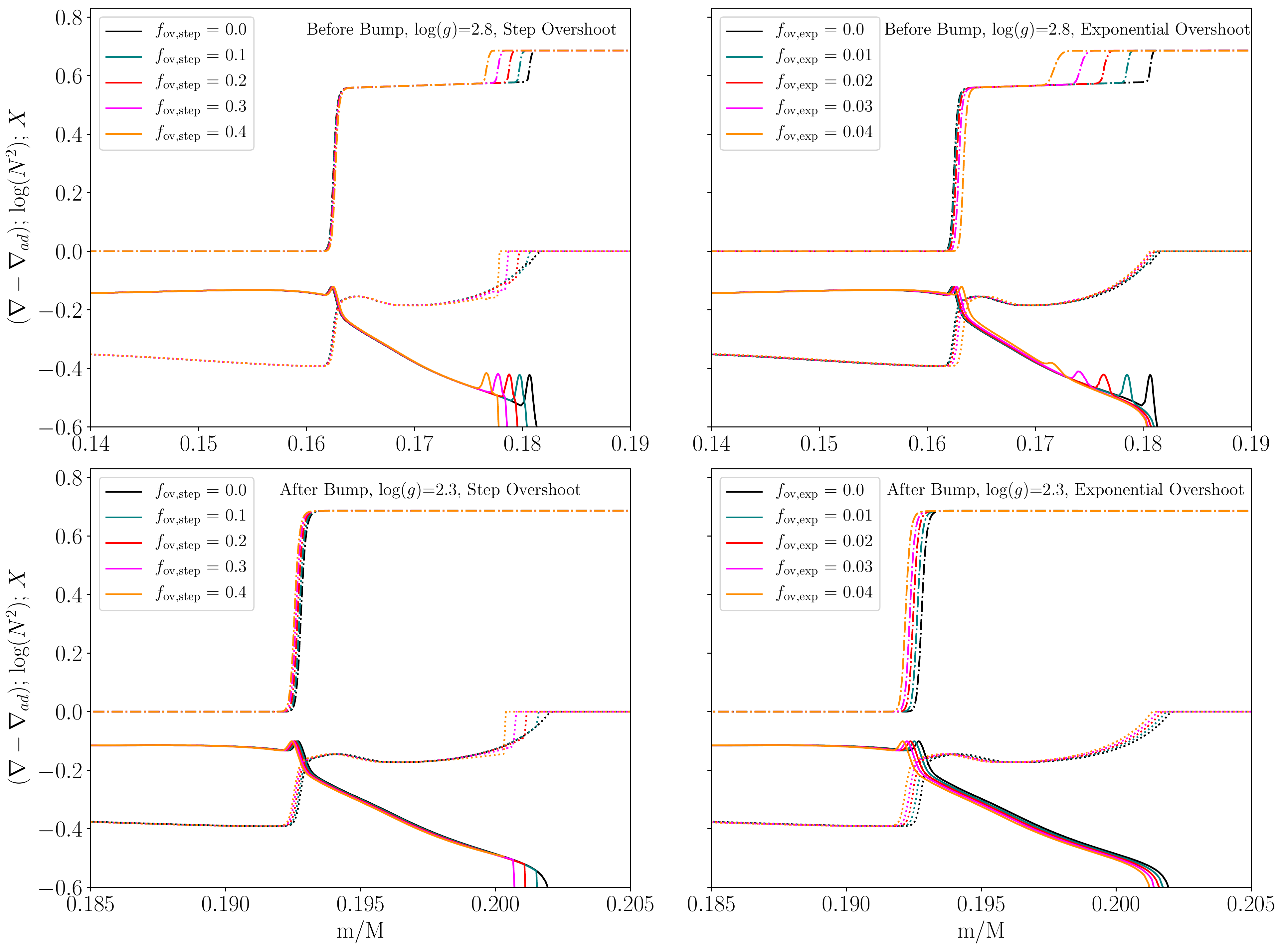}
    \caption{Plots from red giant stellar models showing the profiles in mass coordinate for various important quantities near the convection zone boundary. The dash-dot line shows the Hydrogen abundance ($X$) profile. The solid lines show the natural log of the Brunt–Väisälä frequency profile squared divided by 30. Finally, the dotted lines show the temperature gradient profile minus the adiabatic temperature gradient $\nabla - \nabla_{\text{ad}}$. \\The two plots on the left side show the interior structure for models incorporating penetrative overshoot with a step profile, while the right plots on the right side show the structure for models incorporating exponential profile penetrative overshoot. The upper two plots show the interior structure before the red giant models go through the RGBb and the lower two plots show the structure after the models go through the RGBb.}
    \label{fig:structure_plots}
\end{figure*}

In addition to changing the evolutionary track up the red giant branch, different overshoot prescriptions change the internal structure of our red giant models in varying ways, visualized in \autoref{fig:structure_plots}. Before the red giant stellar model goes through the RGBb, the models' Brunt–Väisälä frequency profiles (solid lines in \autoref{fig:structure_plots}) have glitches (visible as small humps) at the location of the sharp chemical abundance discontinuity (dash-dot lines in \autoref{fig:structure_plots}). The location of these frequency glitches at a given value of log($g$) depends on the amount of overshoot incorporated into the model. 

When considering penetrative overshoot, the depth of the red giant models' convection zone also depends on the amount of overshoot since penetrative overshoot extends the adiabatic temperature gradient, $\nabla_{\text{ad}}$, into the overshoot region, as seen in the temperature gradient profiles (dotted lines in \autoref{fig:structure_plots}). Thus, the amplitude of penetrative overshoot determines the size of the convection region. This is clear from figure \autoref{fig:structure_plots} where we see the Brunt–Väisälä frequency becomes imaginary at deeper depths with increasing overshoot parameter. 

The change in the depth of the convection zone implies that penetrative overshoot prescriptions change the size of the region where g-modes can propagate in the red giant models, thereby changing the mixed-mode properties of the model. In overshoot prescriptions where the temperature gradient of the overshoot region is not changed, (overmixing only, prescriptions 1 and 2) all the dotted lines would match up with the case of no penetrative overshoot.

Once the red giant models have evolved past the RGBb, their abundance profiles match up, since the star's hydrogen-burning shell has passed through the discontinuity in the hydrogen abundance left behind by the first dredge up. Since the abundance profiles match, the small glitches (humps) that interrupt the Brunt–Väisälä frequency profiles before the RGBb are smoothed out. In the case of only overmixing (no difference in temperature gradient between varying overshoot models) there are minimal differences in the oscillation properties for models with different overmixing amplitudes. On the other hand, for models of varying penetrative overshoot, there should still be differences in the oscillation properties even after the RGBb since differing amounts of penetrative overshoot will change the effective size of the g-mode oscillation cavity.

\subsection{The Effects of Overshooting on Asteroseismic Parameters}
For each  model calculated in \autoref{sec:modelling}, we calculated the radial ($\ell = 0$) and dipole ($\ell = 1$) oscillation frequencies using the stellar oscillation code GYRE 6.0 \citep{Townsend2013}. As discussed in \autoref{sec:intro}, pure g-modes or pure non-radial p-modes are not observed in red giants around $\nu_{\text{max}}$. Instead, mixed modes with p-like characteristics in the envelope and g-like characteristics in the core are observed. In red giants, several g-modes couple with each p-mode and the properties of these mixed modes change depending on the implementation of envelope overshoot. 

\subsubsection{Period Spacing}

\begin{figure*}[ht!]
    \centering
    \includegraphics[scale=0.37]{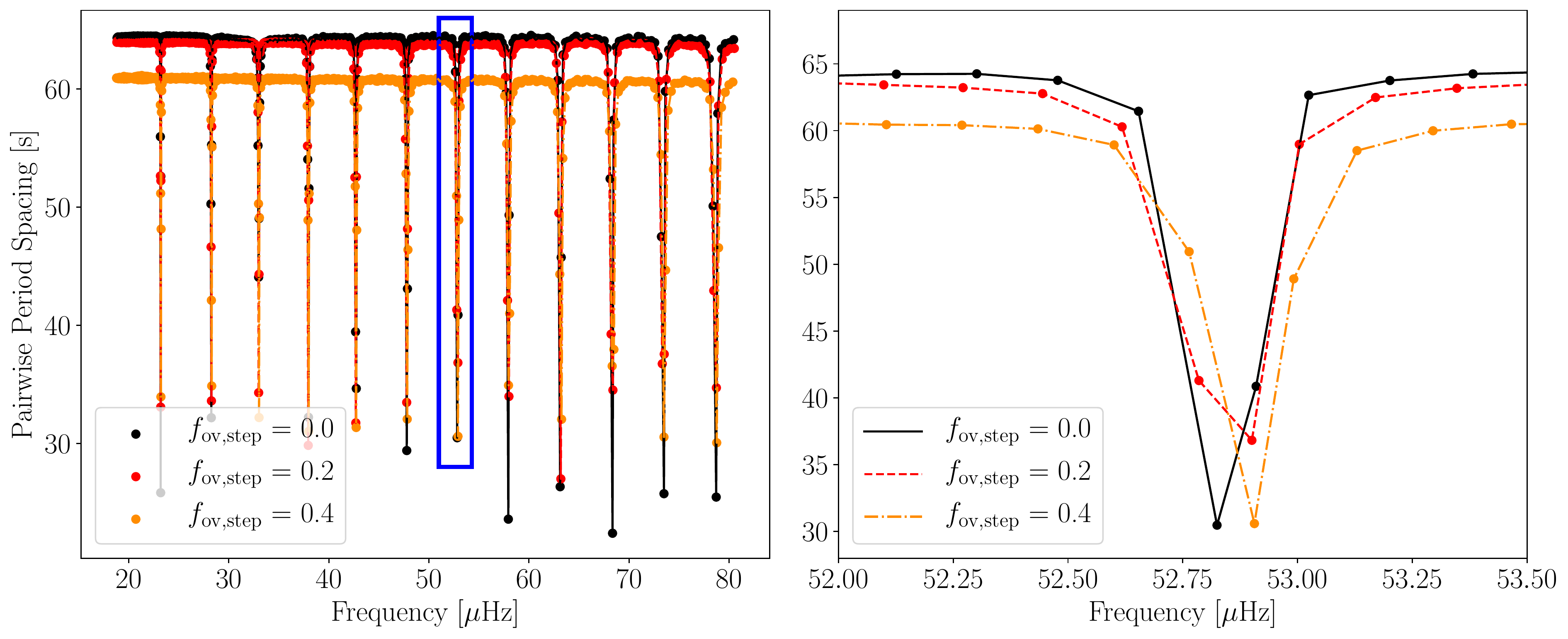}
    \caption{Pairwise period spacing between $\ell = 1$ modes as a function of frequency for three different 1.4$M_{\odot}$ models with solar metallicity and log($g$)=2.6, but with different amounts of step penetrative overshoot. The right-hand panel is a zoom in of the blue box in the left-hand figure. At this value of log($g$), the red giant model with $f_{\text{ov, step}} = 0.4$ has already crossed through the RGBb. However, the other two models are still approaching the RGBb. Note that there is a clear difference between the red giant models before and after the RGBb when you look at the asymptotic period spacing $\Delta\Pi_{1}$ (the modes not part of the downward spikes) but this parameter is generally not observable since only modes with low inertia (bottoms of the peaks) are excited to observable levels. }
    \label{fig:period_spacing}
\end{figure*}

The spacing between mixed mode periods, $\Delta P$, is related to the pure g-mode period spacings, modulated by the strength of the coupling to the p-mode. Note that $\Delta P < \Delta \Pi_{\ell}$, i.e. the integral in \autoref{eq:deltapi} over the red giant's buoyancy cavity (where g-modes can propagate) is smaller than needed to reproduce the observed period spacing \citep{Basu2017AsteroseismicDA,2017A&ARv..25....1H}. In \autoref{fig:period_spacing} we show the pairwise period spacing versus mode frequency for the dipole modes of three red giant models with differing amounts of step envelope overshoot, all at log($g$) = 2.6. The only visible modes are the ones in the peaks with the smallest values of the period spacing; these modes are excited because they have low inertia. The visible modes have small period spacings because they are the most p-dominated mixed modes, interrupting the regular g-dominated pattern. A line connecting the upper modes, not part of the downward peaks, lies approximately at the asymptotic period spacing $\Delta \Pi_{1}$ \citep{Mosser2015}. 

In \autoref{fig:period_spacing} we plot pairwise period spacing against frequency and one can see that models with lower values of step envelope penetrative overshoot (black and red lines) have a higher value of $\Delta \Pi_{1}$ compared with the model with a higher value of overshoot (orange line). This is because the higher overshoot model has evolved past the RGBb by $\log(g)=2.6$ whereas the other, lower amplitude overshoot, models have not. Evolving past the RGBb changes the g-mode period spacing, $\Delta \Pi_{1}$, but since only the lowest period spacing modes are observable, $\Delta \Pi_{1}$ is not generally directly observable in asteroseismic data. The mixed-mode period spacing, $\Delta P$ is observable in real data, but since $\Delta P$ is strongly dependent on the intricate coupling of core g-modes and envelope p-modes, and changes very rapidly with red giant evolution, it is not possible to distinguish between varying overshoot prescriptions directly using the measured $\Delta P$. In practice, some estimator, $\langle\Delta P\rangle$, of the average mixed-mode period spacing is typically constructed out of the available mixed-mode frequencies as an estimator for $\Delta\Pi_{\ell}$.

\begin{figure*}[ht!]
    \centering
    \includegraphics[scale=0.4]{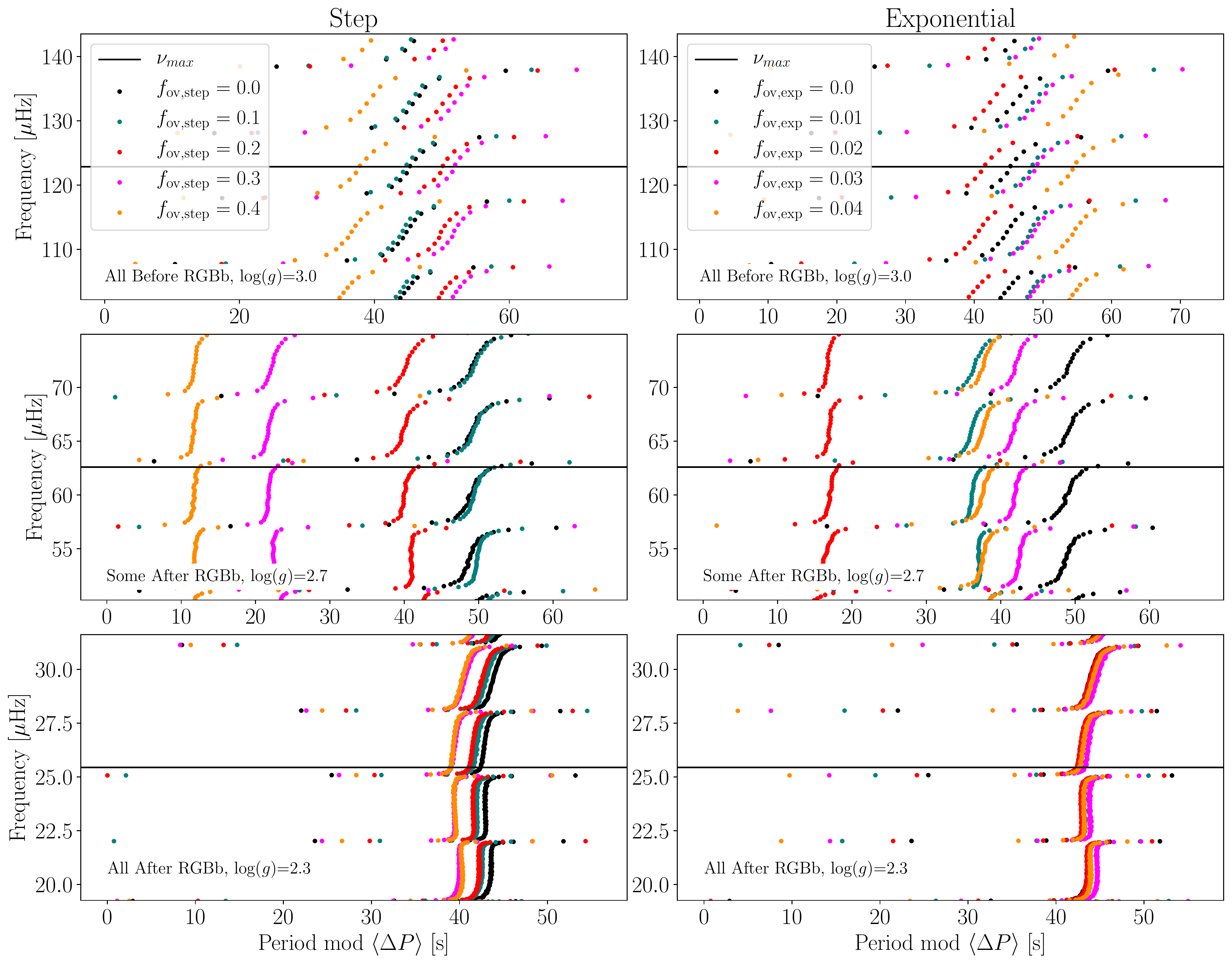}
    \caption{Period \'echelle diagram showing modelled oscillation mode frequencies versus mode period modulo the calibrated average period spacing, $\langle \Delta P \rangle$, for 1.4$M_{\odot}$, solar metallicity red giant models. $\langle \Delta P \rangle$ is found for each model empirically as the period spacing required to straighten out the buoyancy phase function in the neighborhood of $\nu_{\text{max}}$. The left column shows period \'echelle diagrams for models incorporating different amounts of step penetrative overshoot, while the right column shows the results for models incorporating different amounts of exponential penetrative overshoot. The different rows of subplots represent evolutionary states where all the models have yet to enter the RGBb (top row, log($g$)=3.0), where only models with higher $f_{\text{ov, step/exp}}$ have crossed past the RGBb (middle row, log($g$)=2.7), and where all models are past the RGBb (bottom row, log($g$)=2.3). The horizontal black lines show $\nu_{\text{max}}$ for the no overshoot model at the given value of log($g$).   }
    \label{fig:period_echelle_evolution}
\end{figure*}

\subsubsection{Period \'Echelle Diagrams}
An informative way to visualize the asteroseismic differences between red giant models of different overshoot is to plot period \'echelle diagrams comparing models at the same evolutionary value of log($g$). Period \'echelle diagrams show oscillation frequencies versus mode period, modulo the average mixed-mode period spacing, $\langle \Delta P \rangle$. In our subsequent analysis, we determine $\langle \Delta P \rangle$ for each model separately by selecting values such that the s-shaped curves in the corresponding period \'echelle diagrams line up vertically, as depicted in \autoref{fig:period_echelle_evolution}. As the red giant models evolve through the RGBb, the period \'echelle curves of models with varying amounts of overshoot separate as models with greater values of $f_{\text{ov, step/exp}}$ evolve through the RGBb first. In the range of parameter space where $f_{\text{ov, step/exp}}$ determines whether the red giant has evolved past the RGBb, the period \'echelle curves for models of different $f_{\text{ov, step/exp}}$ are easily distinguishable. Once red giants pass through the RGBb, there are no longer major differences in the elemental abundance profiles in the model (see \autoref{fig:structure_plots}) and there is accordingly little separation between models of different $f_{\text{ov, step/exp}}$ in the period \'echelle diagram (bottom panels of \autoref{fig:period_echelle_evolution}).

\begin{figure*}[ht!]
    \centering
    \includegraphics[scale=0.42]{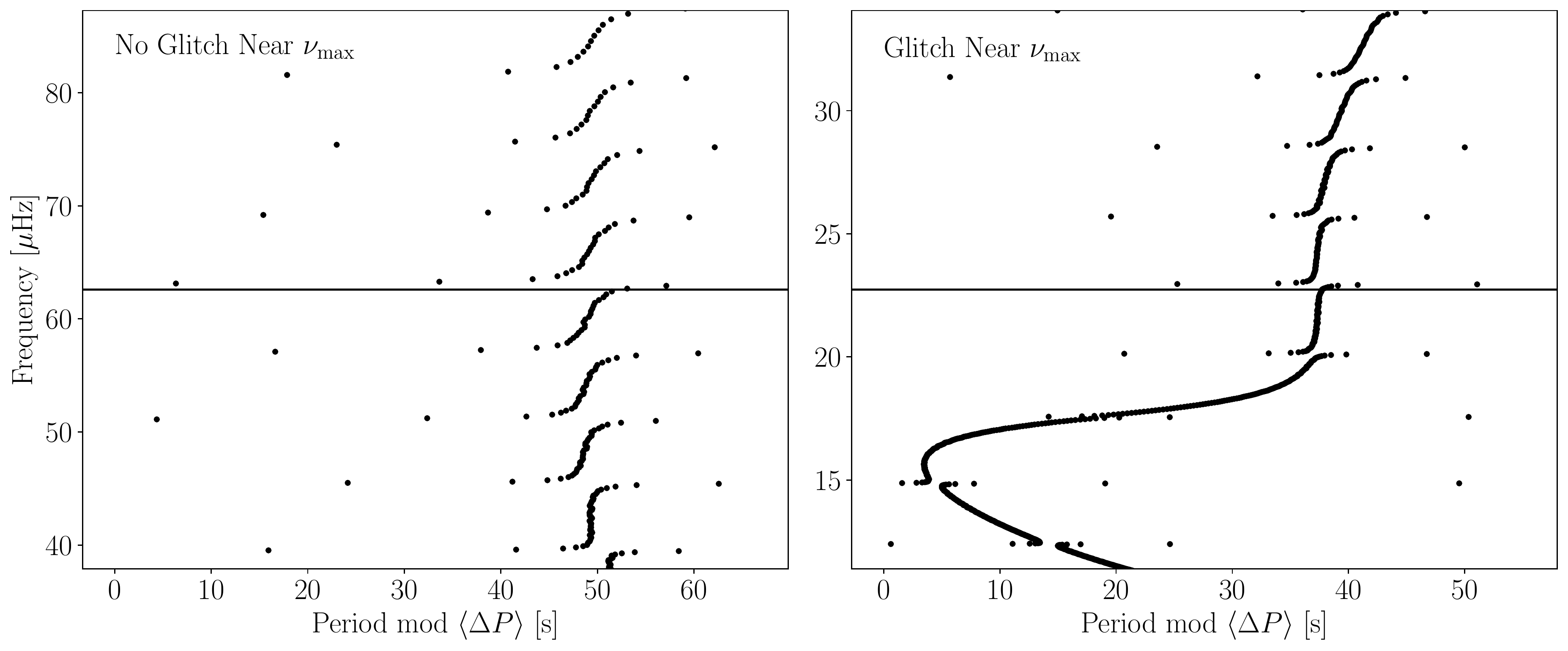}
    \caption{Period \'echelle diagram showing modelled oscillation mode frequencies versus mode period modulo the average mixed-mode period spacing, $\langle \Delta P \rangle$,} for 1.4$M_{\odot}$, solar metallicity red giant models with no overshoot. The panel on the left shows the period \'echelle diagram for the model at $\log(g)=2.7$ before the model goes through the bump, and there is no visible acoustic glitch in the mixed mode period \'echelle pattern near $\nu_{\text{max}}$. The right panel shows the model's period \'echelle diagram once it has reached $\log(g) = 2.25$. At this evolutionary stage, after the model has evolved significantly past the RGBb, there is a large glitch in the period \'echelle pattern near $\nu_{\text{max}}$. 
    \label{fig:Period_echelle_glitch}
\end{figure*}

Glitch signatures, which begin to appear near $\nu_{\text{max}}$ after the red giants evolve past the RGBb, are easily visible in period \'echelle diagrams, as in \autoref{fig:Period_echelle_glitch} where a period \'echelle diagram with no glitch is shown next to a diagram with a glitch. Since the g-mode period spacing is found by aligning the s-shaped curves of the \'echelle diagrams, these glitch signatures make it difficult to determine $\langle \Delta P \rangle$ at low $\log(g)$.

\subsubsection{Gravity-mode Phase Offset}
The position of the vertical period \'echelle curves in \autoref{fig:period_echelle_evolution} change with the evolution of the red giant in a characteristic fashion and is connected to the star's gravity-mode phase offset, $\epsilon_g$, a measure of the phase shift given to the g-modes at the location of the convective boundary.

\begin{figure*}[ht!]
    \centering
    \includegraphics[scale=0.4]{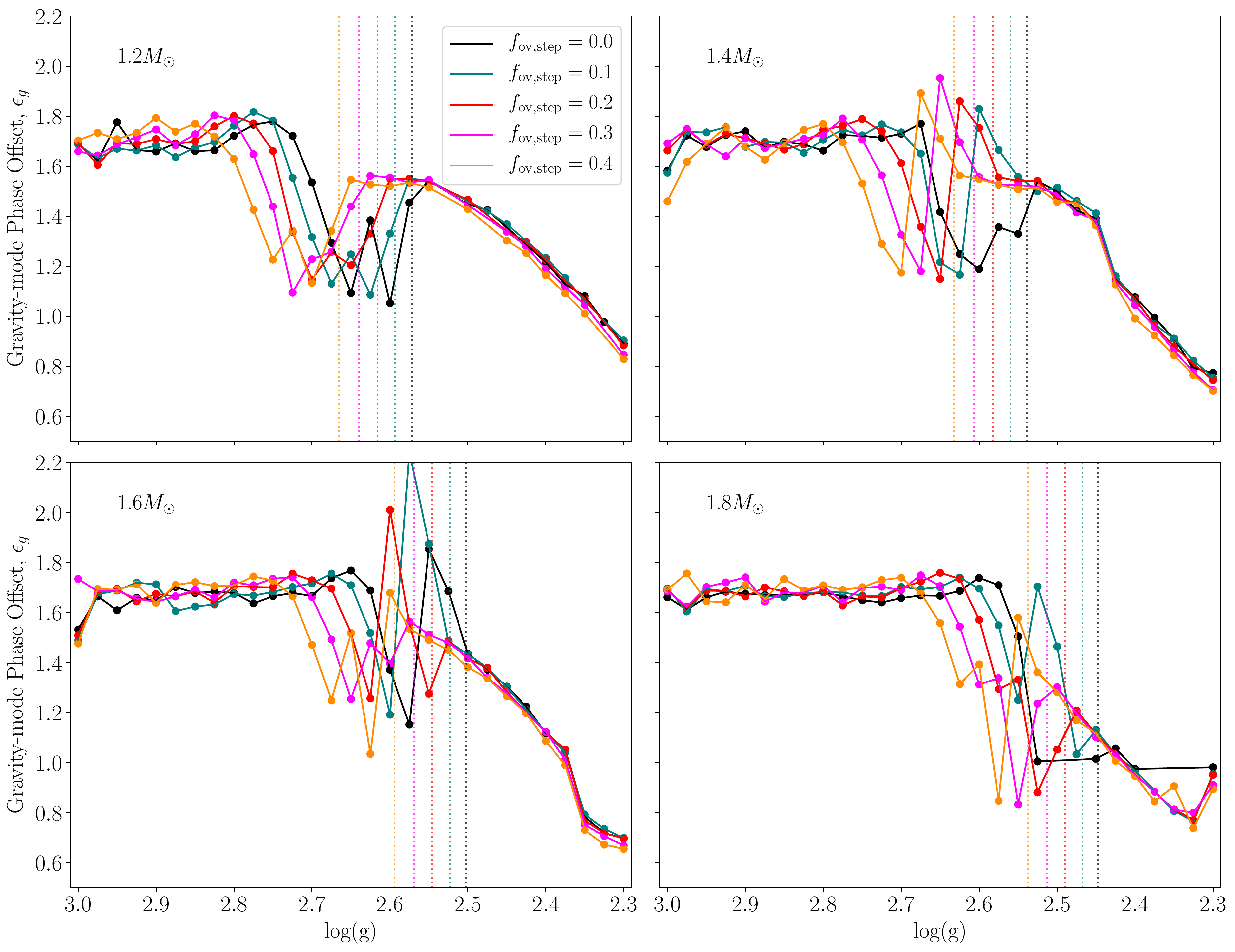}
    \caption{The evolution of the average gravity-mode phase offset, $\epsilon_g$, near $\nu_{\text{max}}$ for solar metallicity models of different masses and with differing amounts of step penetrative overshoot. The different colored lines show the $\epsilon_g$ evolution of models with different values of $f_{\text{ov, step}}$ and the different subplots show $\epsilon_g$ evolution for different model masses. Models evolve from left to right. The vertical lines mark the log($g$) value for the maximum luminosity at the RGBb, where the luminosity starts decreasing. Note that since these $\epsilon_g$ are theoretical values for models, $n_g$ is known for the modes, and we obtain the true offset. For observations, though, $\epsilon_g$ can only be obtained modulo 1.  }
    \label{fig:eg_evolution_masses}
\end{figure*}

For the most g-like mixed modes, with weak coupling to the p-modes (part of the smooth s-shaped curves in \autoref{fig:period_echelle_evolution}), the mode periods can be expressed approximately as the mode periods of pure g-modes, given by
\begin{equation}
    \Pi_{n_g, \ell} = \Delta \Pi_{\ell}(n_g + \epsilon_g(\Pi)),\label{eq:epsg}
\end{equation}
where $\epsilon_g$ is some slowly varying phase function \citep{Aerts-JCD-Kurtz2010}. By equating $\Delta \Pi_{\ell=1}$ to the average mixed-mode period spacing found by aligning the s-shaped curves of the period \'echelle diagram ($\langle \Delta P \rangle$) and by using the $\ell = 1$ mode identification of $n_g$ generated from GYRE, we calculate $\epsilon_g(P)$ via \autoref{eq:epsg} for each mode computed from our red giant models. For each red giant model at each value of log($g$), we then obtain an average value of $\epsilon_g$ by taking the arithmetic mean over the 40 g-like modes closest in frequency to $\nu_{\text{max}}$. Plotting the average $\epsilon_g$ as a function of log($g$) or $\nu_{\text{max}}$ gives the $\epsilon_g$ evolution for our grid of models. At the beginning of our red giant model tracks, near $\log(g) \approx3.5$ the number of modes around $\nu_{\text{max}}$ is small, causing some variation in the average values for $\epsilon_g$, but as the models move up the red giant branch, the number of g-like modes around $\nu_{\text{max}}$ increases greatly so after $\log(g)\approx 3.0$, this method for determining average $\epsilon_g$ is stable.  

\autoref{fig:eg_evolution_masses} shows how different amounts of step penetrative overshoot from the convective envelope changes the $\epsilon_g$ evolution for models with the same mass and metallicity. As predicted in \citet{Pincon_2019}, $\epsilon_g$ appears to remain approximately constant until just before the RGBb when the value decreases smoothly before jumping back up after the RGBb. As the models evolve further past the RGBb, the $\epsilon_g$ value decreases smoothly and rapidly according to our modeling. The $\epsilon_g$ behavior within the RGBb is investigated in higher resolution in \autoref{sec:dis}. Models with different amounts of step $f_{\text{ov, step/exp}}$ evolve past the RGBb at different values of log($g$) so before the RGBb there is a large, clear spread in $\epsilon_g$ values between models dependent on overshoot. After the red giants evolve past the RGBb, there are minimal differences in the average $\epsilon_g$ values for models of different $f_{\text{ov, step/exp}}$ values.

\begin{figure*}[ht!]
    \centering
    \includegraphics[scale=0.5]{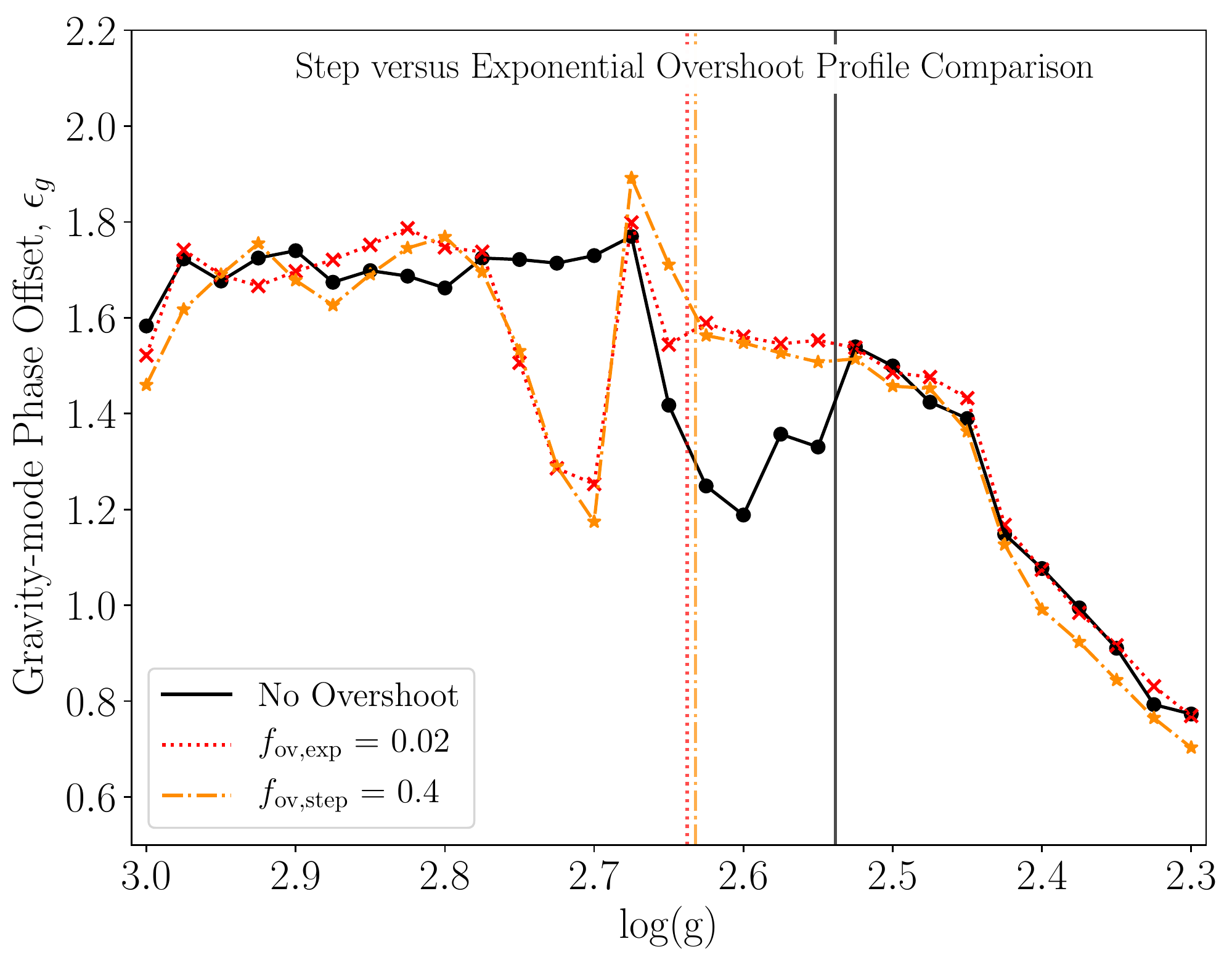}
    \caption{The evolution of the average gravity-mode phase offset, $\epsilon_g$, near $\nu_{\text{max}}$ for solar metallicity models of 1.4 $M_{\odot}$ as a function of the overshoot profile shape. The vertical lines mark the log($g$) value for the maximum luminosity at the RGBb, where luminosity starts decreasing.  }
    \label{fig:eg_evolution_shape}
\end{figure*}

Comparing the $\epsilon_g$ evolution of a red giant model with step penetrative overshoot with the $\epsilon_g$ evolution of a model with exponential penetrative overshoot shows that different overshoot profiles could yield very similar $\epsilon_g$ evolution tracks if the extent of overshoot for the different profile shapes is matched to the position of the RGBb, (i.e. $f_{\text{ov, step}} \approx 20 f_{\text{ov, exp}}$). \autoref{fig:eg_evolution_shape} demonstrates this using a red giant model incorporating exponential penetrative overshoot with $f_{\text{ov, exp}}=0.02$ and another model incorporating step penetrative overshoot with $f_{\text{ov, step}}=0.4$. The $\epsilon_g$ evolution of both models closely match as they evolve through the RGBb. Therefore, \autoref{fig:eg_evolution_shape} demonstrates that matching the $f_{\text{ov, step/exp}}$ value for different overshoot prescriptions by the peak luminosity positions of the RGBb will result in similar $\epsilon_g$ values before, during, and after the RGBb. There is a slight spread in $\epsilon_g$ values between the two models after the models pass the RGBb since the step penetrative overshoot scheme changes the size of the convective ($\nabla=\nabla_{\text{ad}}$) region in a sharper fashion compared with how the exponential penetrative overshoot scheme changes the convective zone boundary (see dotted lines in \autoref{fig:structure_plots}).

\begin{figure*}[ht!]
    \centering
    \includegraphics[scale=0.5]{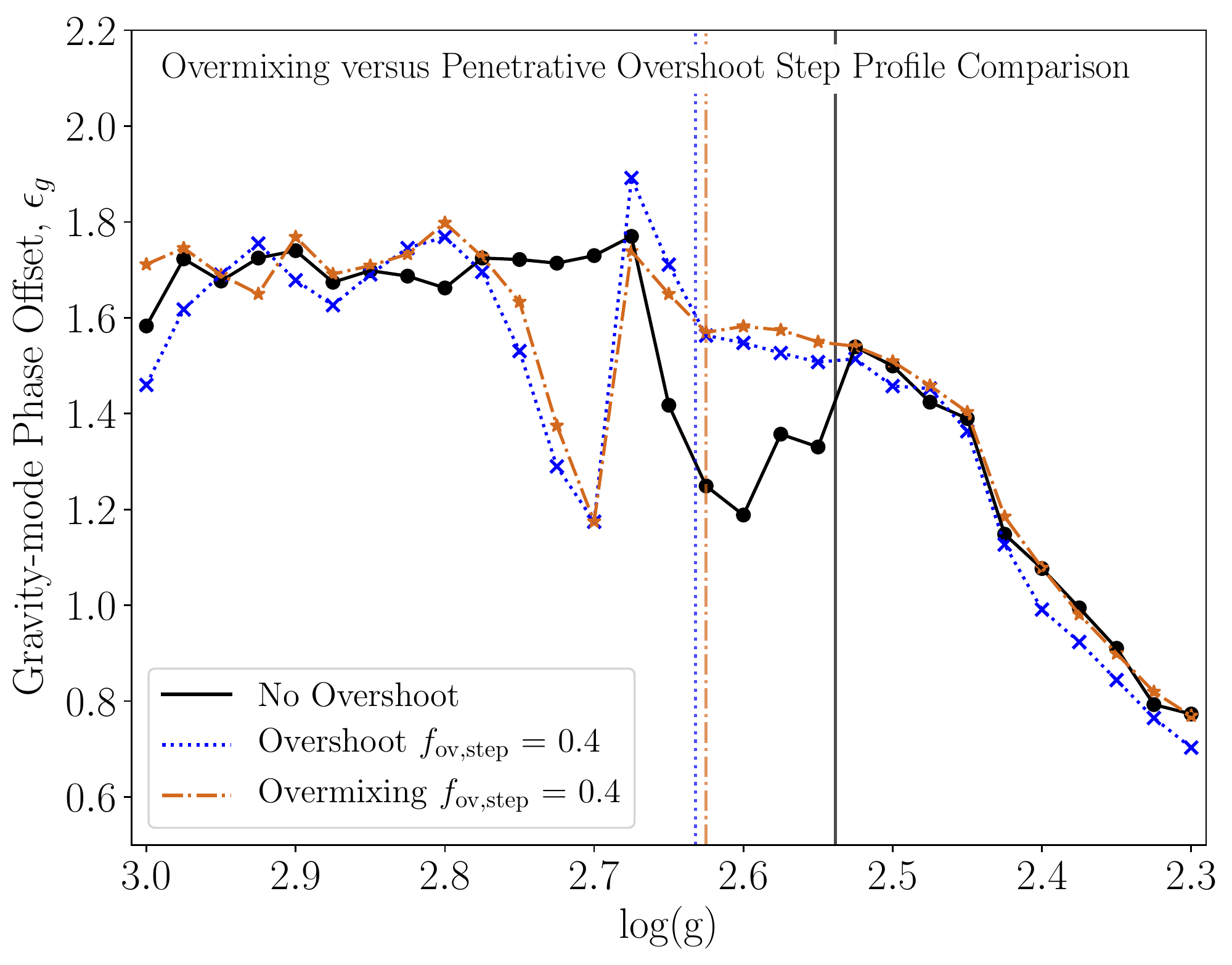}
    \caption{The evolution of the average gravity-mode phase offset, $\epsilon_g$, near $\nu_{\text{max}}$ for solar metallicity models of 1.4 $M_{\odot}$ as a function of the overshoot prescription type. The vertical lines mark the log($g$) value for the maximum luminosity at the RGBb, where the luminosity starts decreasing.}
    \label{fig:eg_evolution_type}
\end{figure*}

Similarly to how we compared the $\epsilon_g$ evolution of our models with different overshoot profile shapes in \autoref{fig:eg_evolution_shape}, we also investigate how modeling overshoot as overmixing or penetrative overshoot can change the $\epsilon_g$ evolution in red giant models where all other parameters are held constant. \autoref{fig:eg_evolution_type} shows that in models where penetrative overshoot is implemented, the modeled $\epsilon_g$ falls from its pre-RGBb value and reaches the top of the RGBb slightly sooner (at higher log($g$)) compared with models with overmixing only. In the penetrative overshoot prescription, the effective depth of the convection zone (where $\nabla = \nabla_{\text{ad}}$) is deepened slightly in comparison to the pure overmixing prescription. Thus, the characteristic drop in $\epsilon_g$ occurs earlier in log($g$).

After the overmixing and penetrative overshoot models evolve past the RGBb, the difference in convection zone depth between the two overshoot treatments causes the $\epsilon_g$ values for the models with penetrative overshoot to be slightly lower when compared to the $\epsilon_g$ models with only overmixing. \autoref{fig:eg_evolution_type} shows the aforementioned effects on $\epsilon_g$ evolution for red giant models incorporating step overmixing and penetrative overshoot with $f_{\text{ov, step}}=0.4$.

\begin{figure*}[ht!]
    \centering
    \includegraphics[scale=0.4]{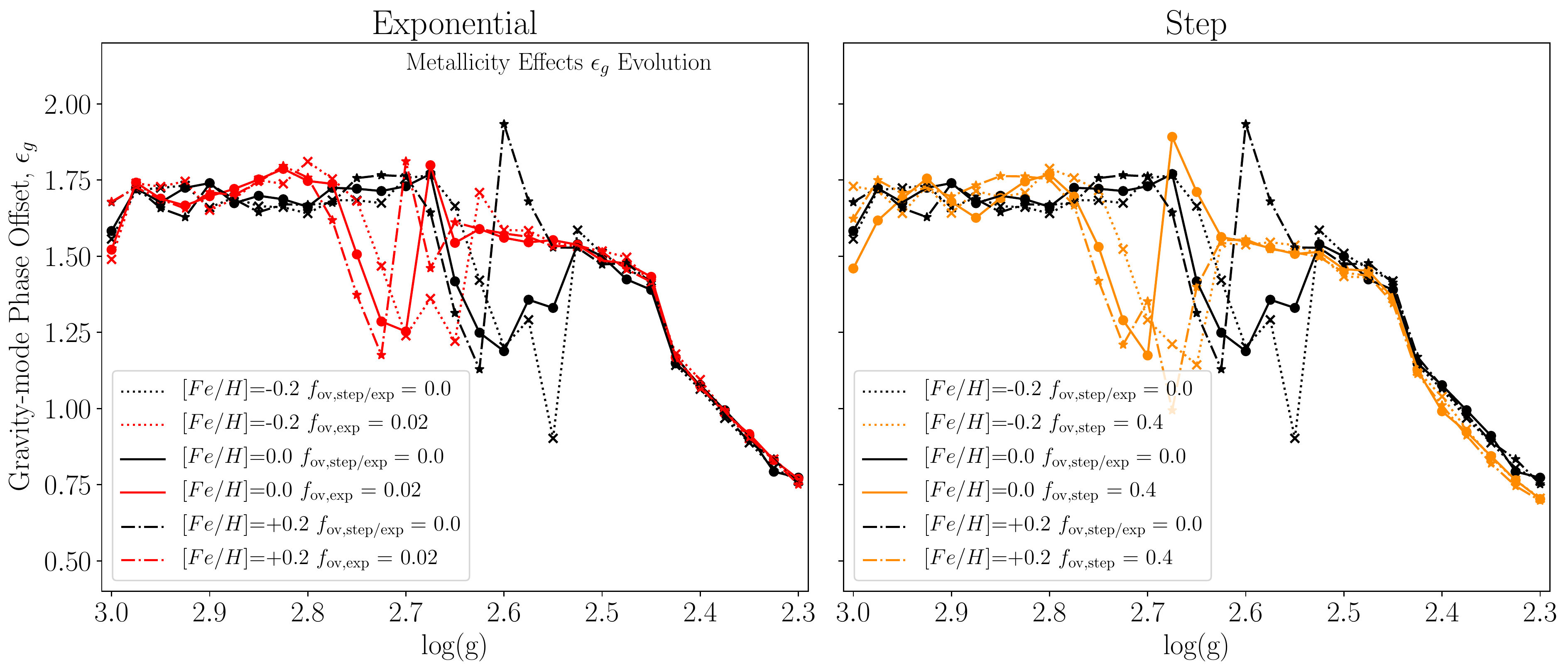}
    \caption{The evolution of the average gravity-mode phase offset, $\epsilon_g$, near $\nu_{\text{max}}$ for solar metallicity models of 1.4 $M_{\odot}$ as a function of the model metallicity. The plot on the left shows the $\epsilon_g$ evolution for models incorporating exponential penetrative overshoot, while the right plot shows the $\epsilon_g$ evolution for models incorporating step penetrative overshoot. }
    \label{fig:eg_evolution_metallicity}
\end{figure*}

As discussed in \autoref{sec:RGBb} and shown in \citet{Khan2018}, the position of the RGBb also depends on the metallicity of the model, in addition to overshoot prescription. Therefore, it is expected that the $\epsilon_g$ evolution of models of different metallicity will also vary. Generally, observational errors on metallicity are around $0.05$ dex, so we change metallicity by more than this to compare how $\epsilon_g$ evolution changes with varying metallicity compared with varying $f_{\text{ov, step/exp}}$. \autoref{fig:eg_evolution_metallicity} shows the $\epsilon_g$ evolution for 1.4$M_{\odot}$ models of varying metallicity and demonstrates that higher metallicity models evolve to the RGBb slightly before lower metallicity models, therefore exhibiting the characteristic drop in $\epsilon_g$ at higher values of log($g$). Models with higher values of $f_{\text{ov, step/exp}}$ also reach the RGBb and experience the characteristic drop in $\epsilon_g$ at higher values of log($g$) compared with models with lower $f_{\text{ov, step/exp}}$. Accurate metallicity measurements to within 0.2 dex are generally available from spectroscopy. \autoref{fig:eg_evolution_metallicity} shows that such variations in metallicity result in changes to $\epsilon_g$ which are less than induced by $\Delta f_\text{ov, exp}$ of 0.01 for exponential overshoot, or $\Delta f_\text{ov, step}$ of 0.2 for step overshoot. These limits on our ability to infer the best-fitting values of $f_\text{ov}$ in real stars are still larger than the values required for consistency with other spectroscopic constraints.

\section{Discussion}

Using a grid of stellar models with varying mass, metallicity, and overshoot, we investigated how the spectroscopic and asteroseismic parameters of red giants respond to different overshoot prescriptions and amounts. In particular, we studied the behavior of the asteroseismic and spectroscopic observable of red giants as they approach and evolve past the red giant branch luminosity bump. Here, we discuss a few other factors. 
\label{sec:dis}
\subsection{Mixing-Length and Initial Helium Abundances}
\begin{figure*}[ht!]
    \centering
    \includegraphics[scale=0.35]{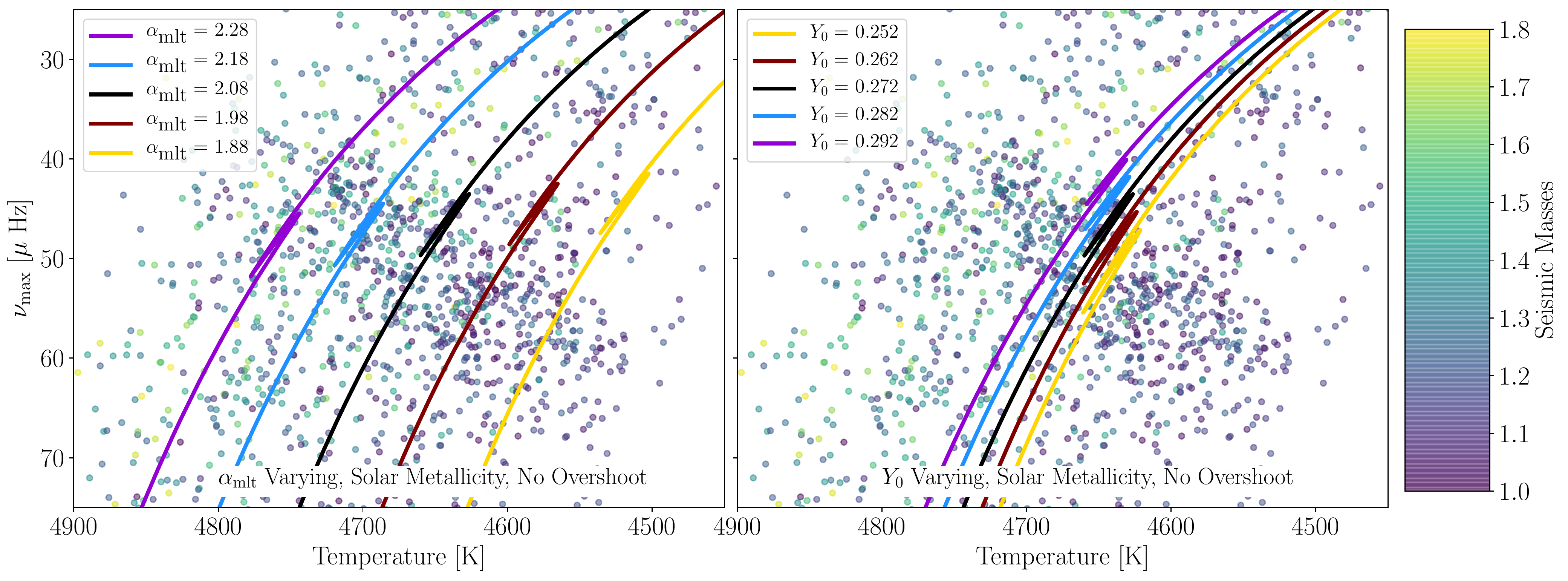}
    \caption{$\nu_{\text{max}}$ versus temperature evolutionary tracks for red giant models incorporating different values of mixing length ($\alpha_{\text{MLT}}$) are shown on the left, plotted along with points taken from the red giant sample of the APOKASC catalog \citep[][]{APOKASC1}. Similarly, evolutionary tracks for red giant models incorporating different values of initial helium abundance ($Y_0$). 
    As the value of $\alpha_{\text{mlt}}$ increases, the $\nu_{\text{max}}$ position at which the model goes through the RGBb also increases (happens earlier). On the other hand, as the value of $Y_0$ increases, the $\nu_{\text{max}}$ position at which the model goes through the RGBb decreases (happens later).  }
    \label{fig:RGBb_alphas}
\end{figure*}

\begin{figure*}[ht!]
    \centering
    \includegraphics[scale=0.5]{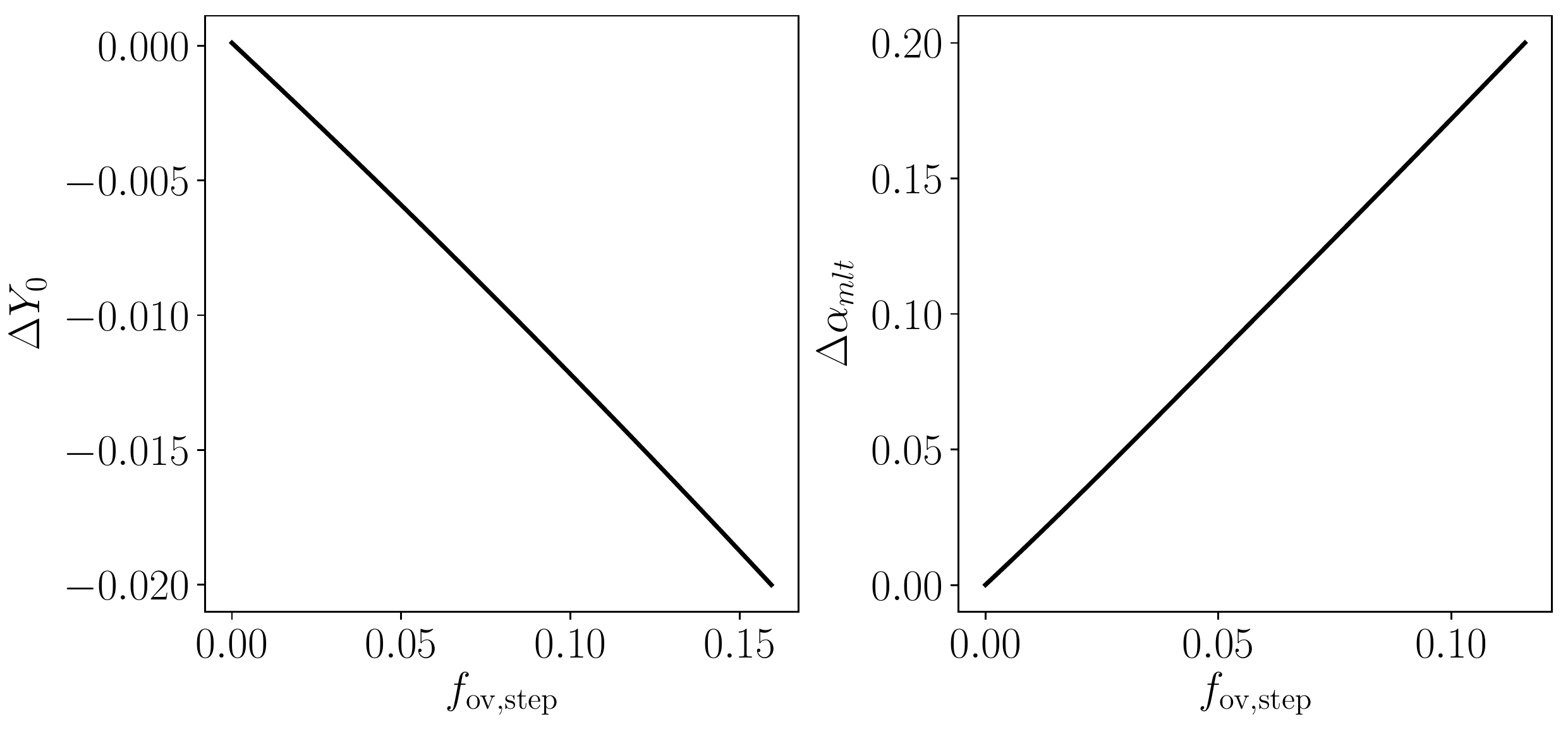}
    \caption{Similar to figure \autoref{fig:f_ov_calibration}, these plots show the calibration of step overshoot parameter $f_\text{ov}$ with changing $Y_0$ (left) and $\alpha_{\text{MLT}}$ (right), using the RGBb as a spectroscopic anchor point. As initial helium abundance decreases from our solar calibrated value of 0.272, the RGBb occurs at higher $\nu_{\text{max}}$ and lower luminosity. As mixing length increases from our calibrated value of 2.08, the RGBb happens at higher $\nu_{\text{max}}$ and lower luminosity. }
    \label{fig:RGBb_helium}
\end{figure*}

There are two other parameters used in stellar models that we have not yet explored here, the initial helium abundance, $Y_0$ and the mixing length parameter, $\amlt$. In this work, we consider $\amlt$ and $Y_0$ in our model grid, assuming them to be equal to the solar calibrated values described in section \autoref{sec:modelling}. However, an increase in $\alpha_{\text{MLT}}$ or decrease in $Y_0$ can lower the position of the RGBb as shown in figure \autoref{fig:RGBb_alphas}, which is degenerate with the effects of overshoot. To illustrate this, we follow the same prescription as in \autoref{fig:f_ov_calibration}, matching up the RGBb positions between evolutionary tracks incorporating different $f_{\text{ov, step/exp}}$, $\alpha_{\text{MLT}}$, and $Y_0$. We find that step overshooting with $f_{\text{ov, step}} = 0.1$ shifts the position of the RGBb about as much as a decrease in $Y_0$ of 0.01 or an increase in $\alpha_{\text{MLT}}$ of 0.15 (\autoref{fig:RGBb_helium}). The position of the RGBb is not the only observable quantity controlled by $Y_0$ and $\amlt$ --- changing these parameters also significantly modifies both the position of the red giant branch (RGB) itself on the HR diagram, as well as the age scale associated with it. In the absence of further constraints, the combination of $Y_0$, $\amlt$, and $f_\text{ov}$ may be tuned arbitrarily to match the observed positions of the RGB and RGBb, in an under-determined fashion --- in this case, with three tuning parameters for two constraints. Additional seismic diagnostics on overshooting are required to break this degeneracy. We have demonstrated that this role may be played by $\epsilon_g$.

\subsection{Observing the Gravity-mode Phase Offset}

\begin{figure*}[ht!]
    \centering
    \includegraphics[scale=0.4]{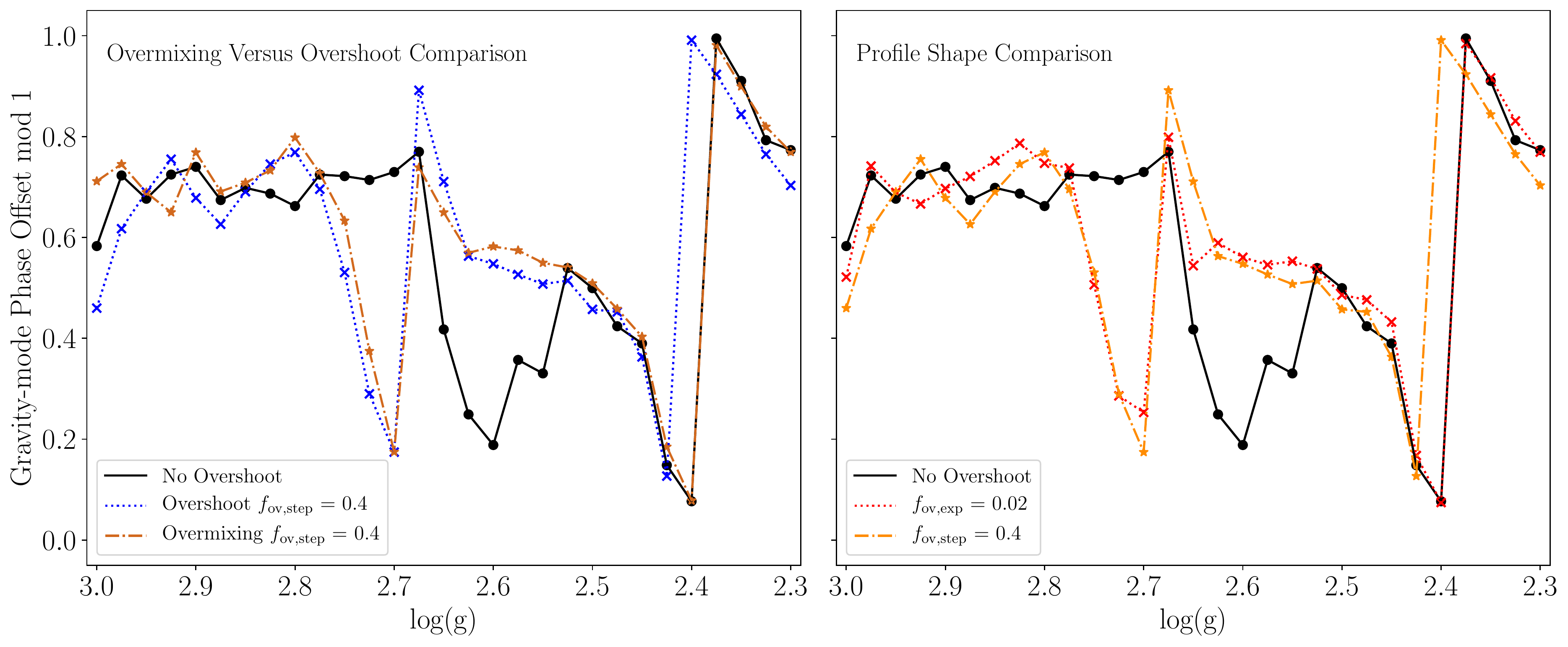}
    \caption{The $\epsilon_g$ values in \autoref{fig:eg_evolution_type} and \autoref{fig:eg_evolution_shape} are theoretical values for models, so we can identify $n_g$ for the modes and obtain the true offset; for observations, $\epsilon_g$ can only be obtained modulo 1. This plot shows the same $\epsilon_g$ evolution as in \autoref{fig:eg_evolution_type} and \autoref{fig:eg_evolution_shape} but taken modulo 1 to mimic what one would see in $\epsilon_g$ evolution plots calculated from real asteroseismic data.}
    \label{fig:eg_evolution_mod_1}
\end{figure*}

In asteroseismic data of red giants, the true radial order $n_g$ is difficult to identify, and it is usually only possible to determine $\epsilon_g$ modulo 1 \citep{Mosser_2018}. Therefore, when using observed asteroseismic parameters to try and distinguish between overshoot prescriptions, one must consider the evolution of $\epsilon_g$ modulo 1, rather than that of its nominal values. The evolution mod 1 is shown in \autoref{fig:eg_evolution_mod_1} for models with varying overshoot region shapes (step versus exponential, right figure) and varying overshoot treatments (overmixing versus overshoot, left figure). \autoref{fig:eg_evolution_mod_1} shows that the general behavior of the gravity mode phase offset evolution is still visible when taking $\epsilon_g$ mod 1. For the most p-dominated mixed modes, $\epsilon_g$ becomes ill-defined. This refers to the modes at the bottom of the period spacing diagram peaks (\autoref{fig:period_spacing}), and the modes far from the S-shaped structures in the period \'echelle diagram (\autoref{fig:period_echelle_evolution}).

Unfortunately, since the most observable modes of oscillating red giants with $\nu_{\text{max}} \lesssim 60 \mu\text{Hz}$ are these most p-dominated modes and, depending on evolutionary stage, gravity dominated mixed modes are difficult to identify, \citep{Dupret2009, Mosser_2018}, using only the p-dominated modes to determine the g-mode period spacing, and hence $\epsilon_g$, presents further methodological difficulties. This may be variously achieved by exploiting the vertical alignment and symmetry of the mixed mode pattern in the period \'echelle diagram \citep{Bedding2011, Datta2015}, or by constructing ``stretched'' \'echelle diagrams \citep{Mosser2015} with some assumptions about the asymptotic behavior of the mixed-mode coupling. The phenomenology we have presented here is intended for consistency with the former construction, pending an analysis of systematic properties of the stretched-\'echelle-diagram construction (Ong, Gehan et al. in prep.). Within the context of this work, however, we note that the asymptotic analysis of \citet{Mosser_2018} and \cite{Pincon_2019}, which underlies the stretched-\'echelle-diagram construction, renders existing literature values for $\epsilon_g$ unequal to those of the true g-modes (or g-dominated mixed modes) returned by numerical calculations as we have used. In particular, existing literature values are offset by 1/2 from --- i.e. exactly out of phase with --- the values associated with the underlying g-modes. While this is a significant correction to make, and essential to our following exposition, it is not the main focus of this paper; we leave further discussion of it to Appendix \ref{appendix}. In our subsequent analysis, we apply an offset to the reported values from \cite{Pincon_2019} and \cite{Mosser_2018}.

\begin{figure*}[ht!]
    \centering
    \includegraphics[scale=0.45]{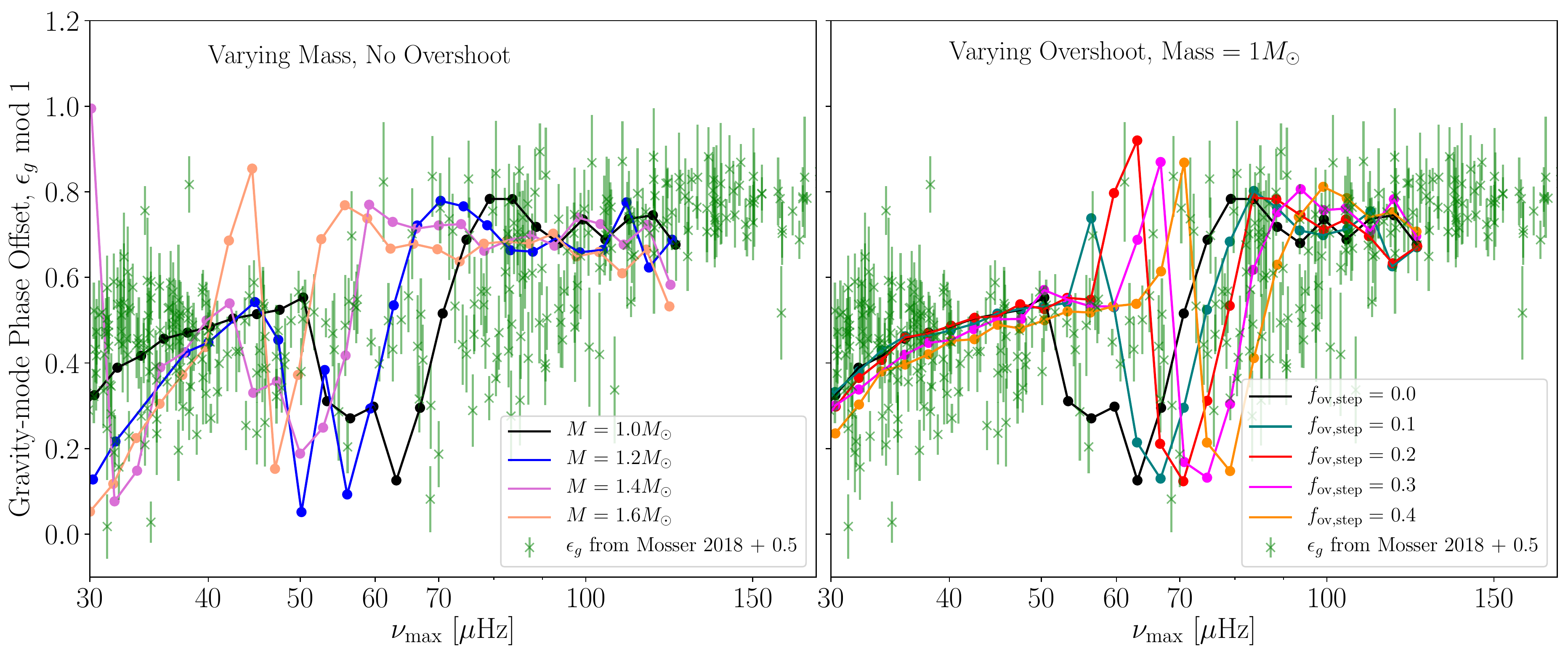}
    \caption{ The gravity mode phase offset, $\epsilon_g$, evolution through the RGBb as a function of $\nu_{\text{max}}$ for our red giant models of varying mass (left panel) and varying overshoot (right panel) are plotted as the solid curves. The green crosses show the $\epsilon_g$ data of \citet{Mosser_2018} plus an offset of one half. }
    \label{fig:Mosser2018comparison}
\end{figure*}

\autoref{fig:Mosser2018comparison} shows the gravity mode phase offset evolution (mod 1) for models of varying mass and $f_{\text{ov, step}}$, plotted along with the $\epsilon_g$ observational data of \citet{Mosser_2018} with the aforementioned offset of one half applied. Comparing the observational $\epsilon_g$ data to the modelled $\epsilon_g$ evolution of different red giant models through the RGBb shows good qualitative agreement before and after the downturn in $\epsilon_g$, which occurs just before the RGBb. With some spread, both the models and the data show that $\epsilon_g$ before the RGBb is around 0.6 to 0.9 while $\epsilon_g$ after the RGBb is lower, around 0.2 to 0.6. The observational data also shows that the decrease in $\epsilon_g$ occurs before (at higher values of $\nu_{\text{max}}$) the decrease seen in the modelled $\epsilon_g$ evolution for models of no overshoot (see left panel of \autoref{fig:Mosser2018comparison}). Envelope overshooting places the modelled downturn in $\epsilon_g$ more in line with the observed downturn in the $\epsilon_g$ data from \citet{Mosser_2018} (see right panel of \autoref{fig:Mosser2018comparison}).   

\subsection{Comparison to Asymptotic Theory}
\citet{Pincon_2019} considered the evolution of the gravity-mode phase offset $\epsilon_g$ using asymptotic analysis in the JWKB approximation \citep{Shibahashi1979,Takata_2016a} and made comparisons to the reported values in \citet{Mosser_2018}. Their analytic model yields reasonable agreement with a portion of the sample of \citet{Mosser_2018}. In particular, their model predicts that as a star evolves up the red giant branch, the value of $\epsilon_g$ will remain approximately constant until just before the RGBb, when the value of $\epsilon_g$ is predicted to smoothly decrease, and tend asymptotically to a different value. In the analysis of \citet{Pincon_2019}, the convective boundary is assumed to be either much further out than, or to coincide with, the outer turning point of the g-mode cavity. In this work, in order to investigate the regime where the convective boundary is approaching the edge of the g-mode cavity, we instead use a grid of red giant models and explicit numerical calculations of g-mode frequencies to study the evolution of the $\epsilon_g$ parameter when stellar properties like overshoot are considered.

\begin{figure*}[ht!]
    \centering
    \includegraphics[scale=0.37]{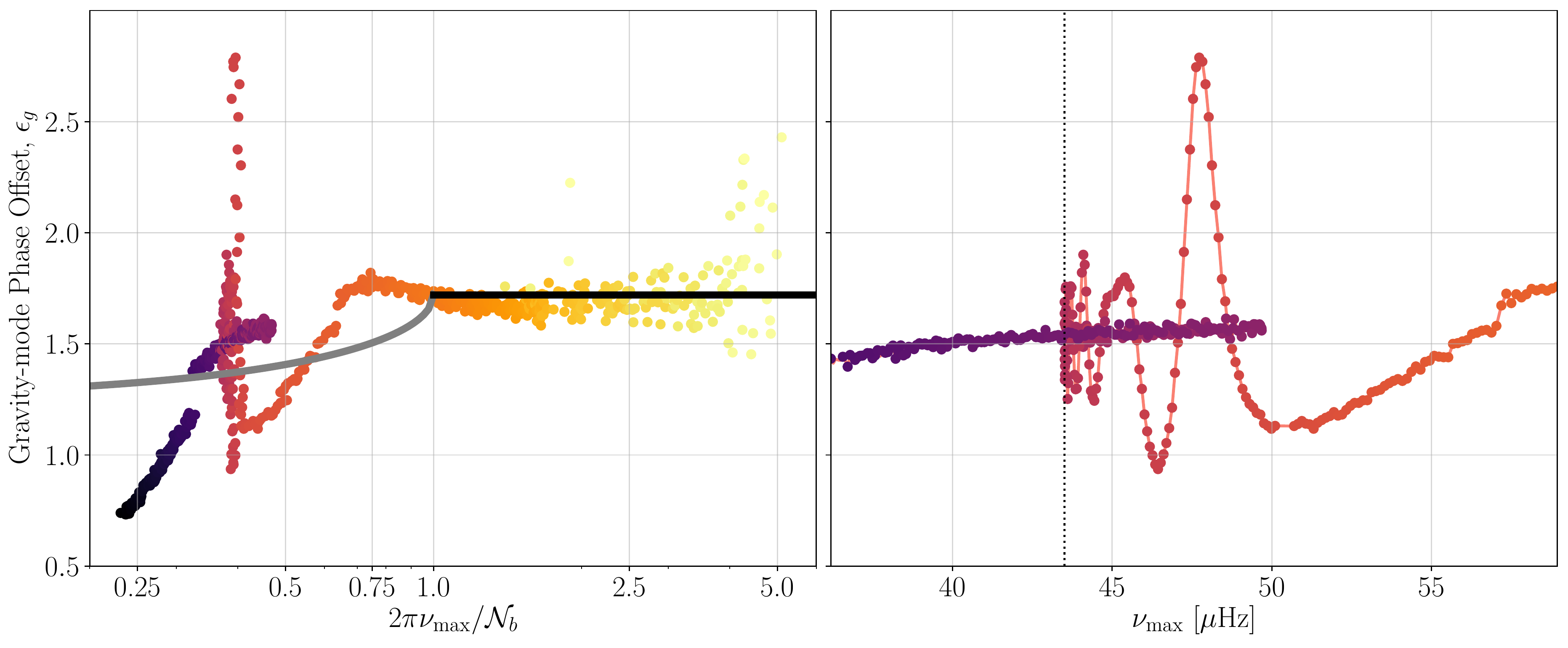}
    \caption{The gravity-mode phase offset, $\epsilon_g$, near $\nu_{\text{max}}$, as a function of $\nu_{\text{max}}$ normalized by the modified Brunt–Väisälä frequency (left panel) and as a function of $\nu_{\text{max}}$ (right panel). The right panel is zoomed in to the region around the RGBb position in $\nu_{\text{max}}$ shown as the vertical line. The points are colored from light to dark according to their increasing age. The solid lines in the left panel show the $\epsilon_g$ evolution model from \citet{Pincon_2019} with the horizontal black line corresponding to their case ``a'' and the gray curved line corresponding to their case ``b''. }
    \label{fig:Pincon_comparison}
\end{figure*}

To illustrate the evolution of $\epsilon_g$ in detail, we took one red giant model track (1.4 $M_{\odot}$, solar metallicity, no overshooting) and calculated its average $\epsilon_g$ using the method in \autoref{sec:asteroseismic} for every time step in its evolution. The left panel of \autoref{fig:Pincon_comparison} plots the $\epsilon_g$ evolution of this red giant model in units of $\nu_{\text{max}}$ normalized by the modified Brunt–Väisälä frequency described in \citet{Pincon_2019}. The solid lines show the analytic model for the evolution of $\epsilon_g$ from \citet{Pincon_2019}. Clearly, as the red giant approaches and crosses the RGBb, its $\epsilon_g$ undergoes significant variation, with fairly complicated behavior which is not captured with the simple asymptotic theory. A clearer view of this evolution through the RGBb is shown in the right panel of \autoref{fig:Pincon_comparison}. Our modelling indicates that the downward turnoff in $\epsilon_g$ predicted by \citet{Pincon_2019} begins earlier in evolution relative to our modeling. We also see that in the lead up to the model passing through the RGBb, there are very significant oscillatory variations in $\epsilon_g$ which are not captured by the asymptotic description. 

These differences are characteristic of the disagreement between the JWKB approximation and our explicit calculations of g-dominated mode frequencies in this phase of evolution. In particular, the distance between the convective boundary and the formal turning point set by the Lamb frequency, is, in this phase of evolution, comparable to the local wavelength of the JWKB wavefunction, and changes relatively quickly. This highly localised structural variation results in rapidly-evolving oscillatory phenomena --- i.e. a glitch, as seen in \autoref{fig:Period_echelle_glitch}. This also suggests that the turnoff in $\epsilon_g$ is caused by a combination of the retreat of the convective boundary, in the lead-up to the RGBb, and the reduction of $\nu_\text{max}$ past the critical buoyancy frequency at the convective boundary, $\mathcal{N}_b/2\pi$, over the course of the star's expansion. Finally, we also see that the evolution of $\epsilon_g$ past the bump does not level off to the constant value suggested by \cite{Pincon_2019}, although this might also be because our models have not yet evolved sufficiently to reenter the regime of accuracy of their asymptotic description.

As apparent in the $\epsilon_g$ data from \citet{Mosser_2018} shown in \autoref{fig:Mosser2018comparison} the significant oscillatory variations in $\epsilon_g$ shown in \autoref{fig:Pincon_comparison} are not observed in this data sample. This is mostly almost certainly due to the fact that this part of evolution through the RGBb, and the concomitant oscillatory variations, is extremely rapid. Another factor is that while plotting the $\epsilon_g$ values derived from our stellar models, we have access to the absolute g-mode radial orders, $n_g$, but we do not have measurements of $n_g$ in actual stars (where measurements of $\epsilon_g$ are made mod 1).

\section{Summary and Conclusion}
\label{sec:conclusion}
We have conducted a theoretical study to investigate how different prescriptions and amounts of convective overshoot from the envelope affects the asteroseismic properties of red giants as they evolve up the red giant branch and through the red giant branch luminosity bump. Our grid of models with varying mass, metallicity, and convective overshoot prescription (detailed in \autoref{sec:modelling}) was created using the 1-D stellar evolution code MESA \citep{Paxton2011, Paxton2013, Paxton2015, Paxton2018, Paxton2019}. The oscillation modes (mixed-mode eigenfrequencies) of stellar models were calculated using the stellar oscillation code GYRE \citep{Townsend2013}. Overshoot is considered as either overmixing or full penetrative overshoot following either a step or exponential profile. 

Plotting the evolution of our stellar models on a seismic HR diagram, ($\nu_{\text{max}}$ versus $T_{\text{eff}}$) we verify the results of \citet{Khan2018} that show that models incorporating higher amounts of overshoot evolve to the red giant branch luminosity bump (RGBb) earlier (at higher values of log($g$) compared with models with lower overshoot).

We investigated the asteroseismic effects of different overshoot prescriptions and amounts and found that as the red giant models evolve past the RGBb, their average g-mode period spacing ($\Delta \Pi_1$) drops significantly. This effect is also apparent when plotting period \'echelle diagrams showing the mixed mode frequency as a function of mode period modulo the period spacing of the g-dominated mixed modes required to straighten out the buoyancy phase function in the neighborhood of $\nu_{\text{max}}$. Our period \'echelle analysis shows that just before the models evolve past the RGBb, models with different amounts of overshoot are well-separated in period \'echelle diagrams, although actually observing these differences will be highly demanding and would require care in determining the period spacing.

We found that another potential diagnostic of overshoot is the evolution of the gravity-mode phase offset parameter, $\epsilon_g$, as a function of log($g$). The average $\epsilon_g$ decreases rapidly before the model evolves past the RGBb, and we also see slight differences between $\epsilon_g$ for models with different overshoot prescriptions (overmixing versus penetrative overshoot or step versus exponential overshoot). 

Using these small differences in $\epsilon_g$ evolution, it would not be possible to discriminate between prescriptions for overshoot when looking at real data without using robust prior estimates of a stellar system's metallicity and $\alpha_{\text{MLT}}$. This prior knowledge can, however, be obtained by studying clusters with \textit{Kepler} data \citep[see][]{McKeever2019}. Currently, the phase offset $\epsilon_g$ is only weakly constrained by \textit{Kepler} observations of red giant mixed mode frequencies \citep{Buysschaert2016}. The determination of $\epsilon_g$ with tight enough confidence intervals to be used in future envelope overshoot studies may therefore only be possible for red giants with a sufficient number of observed mixed modes. For stars on the RGB, with $\nu_{\text{max}} \lesssim 60 \mu\text{Hz}$, \citet{Mosser_2018} showed the difficulty of identifying gravity dominate mixed modes so observing these mixed modes in RGB stars, if possible, will require some caution. Nonetheless, $\epsilon_g$ may still be used to constrain the amount of overshooting for a given prescription, even if the $\epsilon_g$ evolution variations between overshoot prescriptions are too subtle. 

\begin{acknowledgements}
We thank the anonymous referee for their useful comments, which greatly improved this manuscript. We also thank the Yale Center for Research Computing for guidance and use of the research computing infrastructure, as well as W. Ball and M. Cantiello for technical assistance with MESA. CJL was supported by a Gruber Science Fellowship.

\software{
MESA \citep{Paxton2011,Paxton2013,Paxton2015,Paxton2018,Paxton2019}, GYRE \citep{Townsend2013}},
SciPy \citep{scipy}, Pandas \citep{pandas}

The MESA and GYRE inlists we used to generate our models and frequencies, as well as our extra MESA physics hooks are archived on Zenodo and can be downloaded at 
 \dataset[https://doi.org/10.5281/zenodo.7826086]{https://doi.org/10.5281/zenodo.7826086}

\end{acknowledgements}

\bibliographystyle{aasjournal-compact}
\bibliography{ref}

\appendix 

\section{Systematic offsets in literature gravity mode phase offsets}
\label{appendix}

In the JWKB analysis of the mixed-mode problem, the eigenvalue equation for mixed modes is conventionally written in the form
\begin{equation}
    \tan \theta_p(\nu) \cot \theta_g(\nu) = q(\nu),\label{eq:eig}
\end{equation}
(e.g. eq. 1 of \citet{Mosser_2018}), where $q$, $\theta_p$ and $\theta_g$ are frequency-dependent integrals over the stellar structure. Mixed-mode frequencies are obtained from the values of $\nu$ which satisfy this expression. In the case of red giants, these mixed-mode frequencies are derived from a forest of g-modes coupling to a sparse set of p-modes. Accordingly, each mixed-mode frequency can be thought of as depending on an associated g-mode frequency, $\nu_g$, as well as that of the closest p-mode, $\nu_p$. Motivated by this, \cite{Mosser_2018} approximate $\theta_p$ and $\theta_g$ as
\begin{equation}
\theta_p \approx \Theta_p \equiv \pi \left(\nu - \nu_p \over \Delta\nu\right); \theta_g \approx \Theta_g \equiv \pi \left({1 \over \nu} - {1 \over \nu_g} \over \Delta\Pi_l\right).\label{eq:theta}
\end{equation}
This expression for $\theta_g$ is in turn used to derive values of $\epsilon_g$ from stretched \'echelle diagrams in \citet{Mosser_2018} (per their eq. 25), and to translate between JWKB analysis and observational quantities in \citet{Pincon_2019}. However, we show below that it produces values of $\epsilon_g$ which are inconsistent with those associated with the most g-dominated modes (which are closest to the $\nu_g$ underlying these mixed modes).

We note that mixed-mode frequencies may be expanded in powers of the mode coupling strength (in this case $q$) around either $\nu_p$ or $\nu_g$; the roots of \autoref{eq:eig} should therefore tend to either $\nu_p$ or $\nu_g$ as $q \to 0$, following the analysis of \citet{JCD2012,Ong20}, so either $\tan \theta_p$ or $\cot \theta_g$ should vanish in this limit. In particular, this series expansion should converge most rapidly for the most g-dominated mixed modes, as their frequencies are closest to those of the underlying g-modes. As required, $\tan \Theta_p \to 0$ as $\nu \to \nu_p$. However, $\cot \Theta_g = \tan \left({\pi \over 2} - \Theta_g\right)$ is singular as $\nu \to \nu_g$: accordingly, identifying $\theta_g$ with $\Theta_g$ does not allow \autoref{eq:eig} to recover the g-mode eigenvalues in the limit of weak coupling, or for the most g-dominated mixed modes. This inconsistency is resolved by identifying $\cot \theta_g = \tan \left({\pi \over 2} - \theta_g\right) \equiv \tan \Theta_g$ instead, yielding roots of \autoref{eq:eig} at $\nu_g$ as $q \to 0$, as required. We add an additional offset of ${1 \over 2}$ to the reported values of $\epsilon_g$ in both \cite{Mosser_2018} (and associated works), as well as to the theoretical values from \cite{Pincon_2019} for consistency with the g-modes produced from numerical frequency calculations in this work. 


We verify this by comparing our values of $\epsilon_g$ computed with respect to the g-dominated mixed modes of our evolutionary models, as described in \autoref{sec:modelling}, against the values of $\epsilon_g$ from \cite{Mosser_2018} with an additive offset of $1 \over 2$ applied, which we show in \autoref{fig:Mosser2018comparison}. The qualitative agreement between our frequency calculations and the corrected data is discussed in \autoref{sec:dis}.

\end{document}